\documentclass[amssymb,aps,prd,floatfix,nofootinbib,superciptaddress,superscriptaddress]{revtex4}
\usepackage{amsmath}
\usepackage{amssymb}
\usepackage{amsfonts}
\usepackage{graphicx}
\usepackage{mathtools}
\usepackage{mathrsfs}
\usepackage[x11names]{xcolor}
\usepackage[colorlinks=true,linkcolor=blue,citecolor=Blue4,urlcolor=Blue4]{hyperref}
\usepackage{orcidlink}
\usepackage{physics}
\usepackage{tensor}
\usepackage{array}
\usepackage{tabularx}
\usepackage{float}
\usepackage{color}
\usepackage{subcaption}

\renewcommand{\d}{\mathrm{d}}

\def\Opava{Research Centre for Theoretical Physics and Astrophysics, Institute of Physics, Silesian University in Opava, CZ-74601 Opava, Czech Republic}

\begin{document}

\title{Relativistic figures of equilibrium in the Wald magnetosphere}

\author{Pawe{\l} Doruchowski\, \protect\orcidlink{0009-0002-0901-8285}}
\email{pawel.doruchowski@doctoral.uj.edu.pl}
\affiliation{Szko{\l}a Doktorska Nauk \'{S}cis{\l}ych i Przyrodniczych, Uniwersytet Jagiello\'{n}ski}
\affiliation{Instytut Fizyki Teoretycznej, Uniwersytet Jagiello\'{n}ski, {\L}ojasiewicza 11, 30-348 Krak\'{o}w, Poland}

\author{Patryk Mach\, \protect\orcidlink{0000-0003-1493-8668}}
\email{patryk.mach@uj.edu.pl}
\affiliation{Instytut Fizyki Teoretycznej, Uniwersytet Jagiello\'{n}ski, {\L}ojasiewicza 11, 30-348 Krak\'{o}w, Poland}

\author{Audrey Trova\, \protect\orcidlink{0000-0002-4939-0391}}
\email{audrey.trova@ikmail.com}
\affiliation{Center of Applied Space Technology and Microgravity (ZARM), University of Bremen, Am Fallturm 2, 28359 Germany}
\affiliation{Leibniz University of Hannover, Institute for Theoretical Physics, Appelstraße 2, 30167 Hannover, Germany\\}

\author{Bakhtinur Juraev\, \protect\orcidlink{0000-0002-2508-2693}}
\email{bakhtinur.juraev@gmail.com}\affiliation{\Opava}

\begin{abstract}
We consider a self-gravitating, rigidly rotating charged perfect fluid immersed in the Wald magnetosphere, constructed out of two linearly independent Killing vectors present in stationary and axially-symmetric spacetimes. We show that in non-vacuum spacetimes, Wald's solution can be compatible with the electric current associated with a rotating charged perfect fluid characterized by the vanishing electric conductivity. We prove that for rigidly rotating fluids with a constant energy density or described by the polytropic equation of state, the resulting equations expressing the conservation of the energy-momentum tensor can be integrated. Consequently, the system can be described by nearly standard Einstein--Euler equations known from the theory of general-relativistic rotating fluids, with modifications introduced in the Euler--Bernoulli equation. Numerical solutions of the Einstein--Euler equations are provided for these two cases by introducing suitable modifications in the pseudospectral code by Ansorg, Kleinw\"{a}chter, and Meinel.
\end{abstract}

\maketitle

\section{Introduction}

The Wald magnetosphere is an elegant solution to general-relativistic Maxwell equations, valid in stationary and axially symmetric vacuum spacetimes. It was introduced in \cite{Wald1974} by noticing that for a vanishing Ricci tensor, the electromagnetic four-vector potential $A_\mu$ coinciding with a Killing vector constitutes a solution of the Maxwell equations with zero electric current. In a stationary and axially symmetric vacuum spacetime, the Wald magnetosphere can be constructed by assuming the electromagnetic four-vector potential $A_\mu$ in the form of a combination of the timelike and axial Killing vectors (with constant coefficients). For asymptotically flat spacetimes, such a choice corresponds to the electromagnetic field with a uniform asymptotic distribution of the magnetic field. In relativistic astrophysics, this well-known solution serves as a simple and elegant toy-model of magnetospheres around stationary black holes. A very recent account on Wald's construction and other magnetospheres that can be obtained from symmetries---Killing vectors and Killing--Yano tensors---can be found in \cite{Achour2026}.

In this article, we show that, under suitable circumstances, Wald's solution can be generalized to non-vacuum spacetimes. In particular, we consider stationary axially symmetric configurations of rotating magnetized perfect fluids. The numerical examples discussed in this paper are limited to spheroidal (star-like) configurations, but our general discussion can also be applied to perfect fluid tori (with or without black holes). We treat the electromagnetic component as not self-gravitating, though interacting with the fluid dynamically (through the conservation of the total energy momentum tensor). The fluid is assumed to be self-gravitating. The electromagnetic field is still assumed to be given by the vector potential in the form of a combination of available Killing vectors. This results in an electric current associated with the rotating fluid, which has to be taken into account in a consistent way. We show that for rigidly rotating fluids our generalization of Wald's magnetosphere can be consistent with a vanishing electric conductivity (as opposed to the ideal magnetohydrodynamics). In this case, the electric charge density remains ``frozen'' in the fluid.

As particular examples, we consider rigidly rotating configurations of constant energy density and polytropic fluids. Remarkably, the equations expressing the conservation of the energy-momentum tensor can be explicitly integrated in those cases, leading to an equivalent of the Euler--Bernoulli equation. As a consequence, it is possible to obtain numerical solutions of the resulting Einstein--Euler--Maxwell system of equations, by introducing relatively small modifications in existing algorithms designed to compute relativistic figures of equilibrium \cite{Meinel2008}. We give a collection of numerical examples obtained with a modified AKM code written originally by Ansorg, Kleinw\"{a}chter, and Meinel \cite{Ansorg2003}. It is a pseudospectral code, known for providing accurate solutions, and it is freely available as a supplement to Ref.\ \cite{Meinel2008}.

Magnetized self-gravitating rotating stationary perfect fluids were studied from late 1970's, mainly in the context of star-like (spheroidal) systems. In \cite{Bekenstein1978,Bekenstein1979} Bekenstein and Oron considered a possibility to ``dress'' a nonmagnetic neutron star model with a poloidal magnetic field. Subsequent work developed relativistic equilibrium models of magnetized stars with different magnetic field and rotational states \cite{Bocquet1995,Konno1999,Cardall2001,Kiuchi2008,Frieben2012,Ioka2004,Ciolfi2009,Ciolfi2013,Zanotti2002,Bonazzola1993}. Purely poloidal field configurations were studied in \cite{Bocquet1995,Konno1999,Cardall2001,Bonazzola1993}, purely toroidal ones in \cite{Kiuchi2008,Frieben2012}, and mixed poloidal-toroidal ones in \cite{Ioka2004,Ciolfi2009,Ciolfi2013}. Rigid rotation was analyzed in \cite{Bocquet1995,Kiuchi2008,Frieben2012,Bonazzola1993}, whereas \cite{Konno1999,Cardall2001,Ciolfi2009,Ciolfi2013,Ioka2004,Zanotti2002} considered static or slowly rotating models. General, non-circular spacetimes were discussed in \cite{Ioka2003}. A very elegant, formal geometrical analysis of stationary general-relativistic magnetohydrodynamical systems can be found in \cite{Gourgoulhon2011}. Differentially rotating stationary models of neutron stars equipped with both poloidal and toroidal magnetic fields were computed in \cite{Uryu2014}.

Much less is known about magnetized self-gravitating toroidal configurations. Newtonian magnetized self-gravitating disks around a point mass were investigated in \cite{Otani2009}. An analysis of the von Zeipel condition in the context of general-relativistic self-gravitating magnetized disks was presented in \cite{Zanotti2015}. Stationary general-relativistic models of self-gravitating disks with toroidal magnetic fields were constructed in \cite{Mach2019,Dyba2021}, based on the puncture framework developed by Shibata in \cite{Shibata2007}, and assuming a Keplerian rotation prescription introduced in \cite{Karkowski2018a,Karkowski2018b}.

The order of this paper is as follows: In Section \ref{Sec:GenWaldMagn}, we briefly repeat the original construction of Wald's magnetosphere, and extend it to non-vacuum spacetimes associated with a rotating perfect fluid. We derive a corresponding expression for the electric four-current and search for conditions under which this current can be generated by the fluid, in agreement with relativistic Ohm's law. This happens in particular for a rigidly rotating fluid with angular velocity adapted to the electromagnetic field. In Sec.\ \ref{Sec:EulerBernoulli}, we obtain an explicit form of equations expressing the conservation of the energy-momentum tensor (temporarily relaxing the assumption that the Ohm's decomposition remains valid). Subsequently, in Sec.\ \ref{Sec:RigRotFluids}, we specialize to rigidly rotating fluids and solve the resulting conservation equations for configurations with a constant energy density and for the polytropic equation of state. In Sec.\ \ref{Sec:Charge} we discuss the relation between the total electric charge, the charge density, and the Komar expressions for the mass and the angular momentum. Numerical results are presented in Sec.\ \ref{Sec:NumSol}. Finally, Sec.\ \ref{Sec:Finis} provides a summary and conclusions. A more detailed description of the numerical implementation, as well as some technical calculations, are contained in the Appendices.

Throughout this work, we assume the signature of the metric $(-, +, +, +)$ and use the geometric system of units with $c = G = 1$, where $c$ denotes the speed of light and $G$ is the gravitational constant. In addition, we set $\varepsilon_0 = 1/(4\pi)$ for the vacuum permittivity. Then the dimension of the electromagnetic field tensor is that of $(\text{length})^{-1}$, and the electromagnetic potential becomes dimensionless. Spacetime and spatial dimensions are denoted with Greek and Latin indices, respectively.

\section{Generalized Wald's magnetosphere}
\label{Sec:GenWaldMagn}

The classical derivation of Wald's magnetosphere can be summarized as follows \cite{Wald1974}. Let $\xi^\mu$ be a Killing vector. Consider the four-vector potential of the electromagnetic field in the form $A_\mu = \xi_\mu$. Since $\nabla_\mu \xi_\nu + \nabla_\nu \xi_\mu = 0$, the electromagnetic field tensor $F_{\mu\nu} = \partial_\mu A_\nu - \partial_\nu A_\mu$ reads
\begin{equation}
F_{\mu\nu} = \nabla_\mu \xi_\nu - \nabla_\nu \xi_\mu = - 2 \nabla_\nu \xi_\mu.
\end{equation}
Consequently,
\begin{equation}
\nabla_\nu F^{\mu \nu} = - 2 \nabla_\nu \nabla^\nu \xi^\mu.
\end{equation}
Introducing the commutation relation $\nabla_\sigma \nabla_\nu \xi_\mu - \nabla_\nu \nabla_\sigma \xi_\mu = \xi^\lambda R_{\lambda\mu\nu\sigma}$, one can show that
\begin{equation}\label{Killing_id}
    \nabla_\nu \nabla^\nu \xi^\mu = -R\indices{^\mu_\nu} \xi^\nu,
\end{equation}
and thus
\begin{equation}
\label{Maxwell}
\nabla_\nu F^{\mu \nu} = 2 R\indices{^\mu_\nu} \xi^\nu = 4 \pi j^\mu.
\end{equation}
Here $R_{\lambda \mu \nu \sigma}$,  $R\indices{^\mu_\nu}$, and $j^\mu$ denote the Riemann tensor, the Ricci tensor, and the electric current four-vector, respectively.  For vacuum spacetimes $R\indices{^\mu_\nu} = 0$, and the inhomogeneous Maxwell equations \eqref{Maxwell} are satisfied with a vanishing electric current. Note that the sign conventions in the definitions of the Riemann and Ricci tensors used in this paper and in \cite{Wald1974} are different. While this difference does not matter for vacuum spacetimes, it is important in the non-vacuum case.

In the following, we specialize to stationary, axially symmetric and circular metrics, which in cylindrical coordinates $(t,\rho,\varphi,\zeta)$ can be written as
\begin{equation}\label{metricgen}
\d s^2 = g_{tt}(\rho, \zeta) \d t^2 + 2 g_{t\varphi}(\rho, \zeta) \d t \d \varphi + g_{\rho\rho}(\rho, \zeta) \d \rho^2 + g_{\varphi \varphi} (\rho, \zeta) \d\varphi^2 + g_{\zeta\zeta}(\rho, \zeta)\d \zeta^2.
\end{equation}
They admit two linearly independent Killing vectors
\begin{equation}
\label{Killings}
    k^\mu = (1,0,0,0), \quad \eta^\mu = (0,0,1,0).
\end{equation}
In vacuum spacetimes, Wald's magnetosphere is constructed by taking
\begin{equation}\label{Waldpot}
    A^\mu = \xi^\mu = a k^\mu + b \eta^\mu,
\end{equation}
where $a$ and $b$ denote arbitrary constant coefficients. In an asymptotically flat spacetime the electromagnetic potential \eqref{Waldpot} corresponds to a uniform magnetic field of strength $H = 2b$ at infinity (see Appendix \ref{App:EMfields}).

We now ask the question whether the potential \eqref{Waldpot} can remain a valid and consistent solution to the Maxwell equations, even for non-vacuum spacetimes. In particular, we consider the source of the gravitational field in the form of a stationary rotating perfect fluid with the energy-momentum tensor 
\begin{equation}
\label{tmunufluid}
T^\mathrm{(FLUID)}_{\mu \nu} \coloneqq (\varepsilon + p) u_\mu u_\nu +p g_{\mu\nu},
\end{equation}
where $\varepsilon$, $p$, and $u^\mu$ denote the energy density, the pressure, and the components of the four-velocity, respectively. Throughout this paper, we restrict our attention to the case where $u^\rho = u^\zeta = 0$. Thus, the normalization condition of the four-velocity, $u_\mu u^\mu = -1$, reads
\begin{equation}
\label{unorm}
    u_t u^t + u_\varphi u^\varphi = -1. 
\end{equation}
Note that the same can also be written as
\begin{equation}\label{norm_cond}
  g_{tt}(u^t)^2 + 2g_{t\varphi}u^t u^\varphi + g_{\varphi\varphi}(u^\varphi)^2 = -1.
\end{equation}
This allows us to express the time component $u^t$ in terms of the angular velocity $\Omega \coloneqq u^\varphi/u^t$, as
\begin{equation}\label{ut2}
    (u^t)^2 = - \frac{1}{g_{tt} + 2g_{t\varphi}\Omega + g_{\varphi\varphi}\Omega^2}.
\end{equation}

In particular examples, we will consider compact, spheroidal configurations (stars), but our reasoning can also be applied to toroidal configurations (with or without central objects). The Einstein equations corresponding to the source \eqref{tmunufluid} can be written as
\begin{equation}
\label{EinsteinEqs}
R\indices{^\mu_\nu} = 8 \pi \left( {T^\mathrm{(FLUID)}}\indices{^\mu_\nu} - \frac{1}{2}T^\mathrm{(FLUID)} \delta\indices{^\mu_\nu} \right),
\end{equation}
where the trace of the energy-momentum tensor reads
\begin{equation}\label{Tfluid}
    T^\mathrm{(FLUID)} \coloneqq {T^\mathrm{(FLUID)}}\indices{^\mu_\mu} = -\varepsilon + 3p.
\end{equation}
Consequently,
\begin{equation}
R\indices{^\mu_\nu} = 8 \pi \left[ (\varepsilon + p) u^\mu u_\nu + \frac{1}{2}(\varepsilon - p) \delta\indices{^\mu_\nu} \right],
\end{equation}
and, assuming Eq.\ \eqref{Waldpot},
\begin{equation}
    \begin{aligned}
        R\indices{^\mu_\nu} \xi^\nu &= 8 \pi \left[ (\varepsilon + p) u^\mu (a u_\nu k^\nu + b u_\nu \eta^\nu) + \frac{1}{2}(\varepsilon - p) (a k^\mu + b \eta^\mu) \right] \\
        &= 8 \pi \left[ (\varepsilon + p) u^\mu (a u_t + b u_\varphi) + \frac{1}{2}(\varepsilon - p) (a k^\mu + b \eta^\mu) \right].
    \end{aligned}
\end{equation}
For a consistent solution of the Maxwell equations \eqref{Maxwell}, the expression $2 R\indices{^\mu_\nu} \xi^\nu$ has to be interpreted as $4 \pi j^\mu$. This yields the formula for the electric current in the form \begin{equation}
\label{j}
    j^\mu = 4 \left[ (\varepsilon + p) u^\mu (a u_t + b u_\varphi) + \frac{1}{2}(\varepsilon - p) (a k^\mu + b \eta^\mu) \right].
\end{equation}
We emphasize that Eq.\ \eqref{EinsteinEqs} corresponds to the fluid component treated as self-gravitating and the electromagnetic part assumed to be a test field (in the gravitational sense). At the same time, both components will be treated as dynamically coupled---via Maxwell equations \eqref{Maxwell} and the conservation of the total energy-momentum tensor, as described in Sec.\ \ref{Sec:EulerBernoulli}. Neglecting the contribution from the electromagnetic field in the Einstein equations constitutes an approximation, and it has consequences regarding consistency between conservation of the energy-momentum tensor and the Einstein equations. We will return to this point in the forthcoming discussion.

In general-relativistic magnetohydrodynamics, the electric current four-vector is often orthogonally decomposed with respect to the four-velocity $u^\mu$ as
\begin{equation}\label{j_decomp}
    j^\mu = \rho_\mathrm{c} u^\mu + \sigma F^{\mu \nu} u_\nu,
\end{equation}
where $\rho_\mathrm{c} \coloneqq - j^\mu u_\mu$ denotes the density of the electric charge in the frame comoving with the fluid, and $\sigma$ is the electric conductivity (see, e.g., \cite{Tsagas2005}). This decomposition is usually referred to as relativistic Ohm's law. Note that the orthogonal part of $j^\mu$, $\sigma F^{\mu \nu} u_\nu$, is a spacelike vector or zero. We now check the conditions under which our model remains compatible with the above form. For the electromagnetic potential as in Eq.\ \eqref{Waldpot}, one has
\begin{equation}\label{A_cov}
    A_\mu = ak_\mu + b\eta_\mu = ag_{\mu t} + bg_{\mu\varphi},
\end{equation}
and then
\begin{equation}\label{F0}
    F_{t\mu} u^\mu = F_{tt}u^t + F_{t\varphi}u^\varphi = 0, \qquad F_{\varphi\mu} u^\mu = F_{\varphi t}u^t + F_{\varphi\varphi}u^\varphi = 0
\end{equation}
(recall that $u^\rho = u^\zeta = 0$). Consequently, the part $\sigma F^{\mu \nu} u_\nu$ remains orthogonal not only to $u^\mu$, but also to the Killing vectors \eqref{Killings}, and hence to $j^\mu$ given by Eq.\ \eqref{j}. As such, $\sigma F^{\mu \nu} u_\nu = j^\mu - \rho_\mathrm{c}u^\mu$ must be orthogonal to itself, which is only possible if it vanishes, i.e., if
\begin{equation}\label{j_decomp_sigma0}
    j^\mu = \rho_\mathrm{c} u^\mu.
\end{equation}
A posteriori, we check that for solutions derived in this paper the term $F^{\mu\nu}u_\nu$ does not vanish, and thus $\sigma \equiv 0$. We conclude that the fluid remains in the regime opposite to the ideal magnetohydrodynamics, i.e., it behaves like a perfect insulator and not a perfect conductor. Non self-gravitating equilibrium configurations of charged fluid in such regime have been extensively studied. Indeed, starting with \cite{Kovar11} and later on with \cite{Trova2020,2024aTrova}, equipotential surfaces of charged fluid surrounding a (un)static black hole immersed in an asymptotically homogeneous magnetic field described by Wald’s test-field solution of the Maxwell equations have been built. The closed equipotential surfaces exhibit various geometries, such as single or double tori in the equatorial plane, levitating lobes or polar clouds. Their shape depends highly on the strength of external magnetic field and the charge of the fluid.

Relation \eqref{j_decomp_sigma0}, in turn, imposes additional constraints. Comparing it with Eq.\ \eqref{j}, we get
\begin{subequations}
\begin{align}
    \rho_\mathrm{c} u^t &= 4 \left[ (\varepsilon + p) u^t (a u_t + b u_\varphi) + \frac{a}{2}(\varepsilon - p) \right], \label{sigmazeroa} \\
    \rho_\mathrm{c} u^\varphi &= 4 \left[ (\varepsilon + p) u^\varphi (a u_t + b u_\varphi) + \frac{b}{2}(\varepsilon - p) \right]. \label{sigmazerob}
\end{align}
\end{subequations}
Multiplying the first of the above equations by $u^\varphi$, the second one by $u^t$, and subtracting the results, we come to
\begin{equation}
(\varepsilon - p) a u^\varphi = (\varepsilon - p) b u^t.
\end{equation}
Hence, compatibility with the decomposition \eqref{j_decomp} implies that either $\varepsilon = p$ or $\Omega = b/a = \mathrm{const}$. The first case leads to the so-called ultra-hard equation of state (see, e.g., \cite{Petrich1988}). The second possibility amounts to rigid rotation. Setting $\Omega = b/a$, one obtains
\begin{equation}\label{rhoc_rigid}
    \rho_\mathrm{c} = 4(\varepsilon + p) a \left[ u_t + \Omega u_\varphi + \frac{\varepsilon - p}{2(\varepsilon + p) u^t} \right] = -\frac{2 a}{u^t}(\varepsilon + 3 p),
\end{equation}
where in the last step we have used the normalization \eqref{unorm}.

For later use, let us also note that if $\Omega = b/a$, then
\begin{equation}
    F_{\mu\nu}u^\nu = F_{\mu t}u^t + F_{\mu\varphi}u^\varphi = u^t(\partial_\mu A_t + \Omega\partial_\mu A_\varphi) = au^t\partial_\mu(g_{tt} + 2\Omega g_{t\varphi} + \Omega^2g_{\varphi\varphi}),
\end{equation}
where we have employed Eq.\ \eqref{A_cov}. Since $g_{tt} + 2\Omega g_{t\varphi} + \Omega^2g_{\varphi\varphi} = -(u^t)^{-2}$ in view of identity \eqref{ut2}, we get the relation
\begin{equation}\label{sigma0}
    F_{\mu\nu}u^\nu = 2a\frac{\partial_\mu u^t}{(u^t)^2} = -2a\partial_\mu\left(\frac{1}{u^t}\right).
\end{equation}
In other words, the electric field measured by an observer comoving with the fluid (see Appendix \ref{App:EMfields}) is proportional to the gradient of $1/u^t$.

In the following, we will first consider a more general case, not necessarily satisfying Eq.\ \eqref{j_decomp}, and then assume rigid rotation with $\Omega = b/a$ and $j^\mu = \rho_\mathrm{c} u^\mu$.

\section{Conservation of the energy-momentum tensor}
\label{Sec:EulerBernoulli}

We take into account the back-reaction of the electromagnetic field on the fluid by assuming the conservation of the energy-momentum tensor of the form
\begin{equation}
\label{EulerBernoulli}
\nabla_\mu T\indices{^\mu_\nu} = 0, \qquad T\indices{^\mu_\nu} \coloneqq {T^\mathrm{(FLUID)}}\indices{^\mu_\nu} + {T^\mathrm{(EM)}}\indices{^\mu_\nu},
\end{equation}
where
\begin{equation}
{T_\mathrm{(EM)}}\indices{^\mu^\nu} \coloneqq \frac{1}{4 \pi} \left( F\indices{^\mu_\alpha} F^{\nu \alpha} - \frac{1}{4} F_{\alpha \beta} F^{\alpha \beta} g^{\mu \nu} \right)
\end{equation}
denotes the energy-momentum tensor of the electromagnetic field.

A standard calculation shows that
\begin{equation}
\label{divT_fin}
    \nabla_\mu {T^\mathrm{(FLUID)}}\indices{^\mu_\nu} = \partial_\nu p - (\varepsilon + p)\left(\frac{\partial_\nu u^t}{u^t} - u^t u_\varphi\partial_\nu\Omega\right),
\end{equation}
where $\Omega = u^\varphi/u^t$ (not necessarily constant). For completeness, the derivation is provided in Appendix \ref{App:Bernoulli}.

The electromagnetic term in Eq.\ \eqref{EulerBernoulli} can be treated as follows. One has
\begin{equation}
\label{BernoulliEM}
\nabla_\mu {T_\mathrm{(EM)}}\indices{^\mu_\nu} = j^\mu F_{\mu \nu} = - j^\mu \partial_\nu A_\mu,
\end{equation} 
where we have assumed that $j^\rho = j^\zeta = 0$ (this will automatically be true if relation \eqref{j_decomp_sigma0} is satisfied). Inserting the expression for the electric current \eqref{j} into Eq.\ \eqref{BernoulliEM}, we get
\begin{equation}\label{divTEM}
\nabla_\mu {T_\mathrm{(EM)}}\indices{^\mu_\nu} = -4 \left[ (\varepsilon + p) u^\mu (a u_t + b u_\varphi) + \frac{1}{2}(\varepsilon - p) (a k^\mu + b \eta^\mu) \right] \partial_\nu A_\mu. 
\end{equation}
Taking Eq.\ \eqref{A_cov}, as well as the normalization condition \eqref{unorm}, into account, we find
\begin{equation}\label{divTEM1}
    \begin{aligned}
        u^\mu(au_t + bu_\varphi)\partial_\nu A_\mu &= u^t(au_t + bu_\varphi)(\partial_\nu A_t + \Omega\partial_\nu A_\varphi) \\
        &= \left[-a(1 + u_\varphi u^\varphi) + bu_\varphi u^t\right]\left[\partial_\nu(A_t +\Omega A_\varphi) - A_\varphi\partial_\nu\Omega\right] \\
        &= \left[-a + u_\varphi u^t(b - a\Omega)\right]\left[\partial_\nu(ag_{tt} + bg_{t\varphi} + a\Omega g_{t\varphi} + b\Omega g_{\varphi\varphi}) - (ag_{t\varphi} + bg_{\varphi\varphi})\partial_\nu\Omega\right],
    \end{aligned}
\end{equation}
and
\begin{equation}\label{divTEM2}
        (ak^\mu + b\eta^\mu)\partial_\nu A_\mu = a\partial_\nu A_t + b\partial_\nu A_\varphi = \partial_\nu(aA_t + bA_\varphi) = \partial_\nu(a^2g_{tt} + 2abg_{t\varphi} + b^2g_{\varphi\varphi}).
\end{equation}
Substituting Eqs.\ \eqref{divTEM1} and \eqref{divTEM2} into Eq.\ \eqref{divTEM}, we finally obtain the formula
\begin{equation}
    \begin{aligned}
        \nabla_\mu {T_\mathrm{(EM)}}\indices{^\mu_\nu} = {} & -4(\varepsilon + p)\left[-a + u_\varphi u^t(b - a\Omega)\right]\left[\partial_\nu(ag_{tt} + bg_{t\varphi} + a\Omega g_{t\varphi} + b\Omega g_{\varphi\varphi}) - (ag_{t\varphi} + bg_{\varphi\varphi})\partial_\nu\Omega\right] \\
        & - 2(\varepsilon - p)\partial_\nu(a^2g_{tt} + 2abg_{t\varphi} + b^2g_{\varphi\varphi}). \label{divTEM_fin}
    \end{aligned}
\end{equation}

In principle, one can now study integrability conditions resulting from Eqs.\ \eqref{EulerBernoulli}, \eqref{divT_fin}, and \eqref{divTEM_fin}.  In the remainder of this paper we show that these equations can be integrated for a particularly simple choice of a rigidly rotating fluid with $\Omega = \mathrm{const} = b/a$.

\section{Rigidly rotating fluids}
\label{Sec:RigRotFluids}

In the following, we will discuss the equations expressing the conservation of the total energy-momentum tensor (leading to an analog of the Euler--Bernoulli equation) for rigidly rotating fluids. The set of equations required to describe the system can be closed by (some of) the Einstein equations \eqref{EinsteinEqs} written for the metric \eqref{metricgen}. We emphasize that our construction is based on an approximation---we consider both components of the energy-momentum tensor, $T_\mathrm{(FLUID)}^{\mu \nu}$ and $T_\mathrm{(EM)}^{\mu \nu}$, in the conservation law \eqref{EulerBernoulli}, but only take into account the fluid part $T_\mathrm{(FLUID)}^{\mu \nu}$ in the Einstein equations. Consequently, the latter cannot be compatible in the strict sense with the Euler--Bernoulli equation involving both components of the energy-momentum tensor (Eqs. \eqref{Heqconstsol} and \eqref{HeqsolPoly} below). In practice, we only take into account 4 of the Einstein equations, adhering to a standard formulation known from the theory of stationary axially-symmetric, self-gravitating perfect fluids (see Sec.\ \ref{Sec:NumSol}).

For rigidly rotating fluids with $\Omega = b/a$, Eq.\ \eqref{divTEM_fin} reduces to
\begin{equation}
    \nabla_\mu {T_\mathrm{(EM)}}\indices{^\mu_\nu}  = 2 (\varepsilon + 3p) \partial_\nu(a^2g_{tt} + 2abg_{t\varphi} + b^2g_{\varphi\varphi}).
\end{equation}
On the other hand, identity \eqref{ut2} implies that
\begin{equation}\label{ut_identity}
    a^2g_{tt} + 2abg_{t\varphi} + b^2g_{\varphi\varphi} = a^2 (g_{tt} + 2\Omega g_{t\varphi} + \Omega^2g_{\varphi\varphi}) = - \frac{a^2}{(u^t)^2}.
\end{equation}
Hence,
\begin{equation}\label{divT2}
    \nabla_\mu {T_\mathrm{(EM)}}\indices{^\mu_\nu}  = -2 a^2 (\varepsilon + 3p) \partial_\nu \left[ \frac{1}{(u^t)^2} \right].
\end{equation}
Note that this result can also be obtained by noticing that for $\Omega = b/a$, the electric current four-vector has the form
\begin{equation}
    j^\mu = \rho_\mathrm{c} u^\mu = -2 a (\varepsilon + 3p) \frac{u^\mu}{u^t},
\end{equation}
as follows from Eq.\ \eqref{rhoc_rigid}, and using directly formula \eqref{BernoulliEM}.

Combining Eqs.\ \eqref{divT_fin} and \eqref{divT2}, one gets
\begin{equation}\label{E-B_eq_rigid3}
    \partial_\nu p - (\varepsilon + p) \partial_\nu(\ln u^t) - 2 a^2 \left( \varepsilon + 3 p \right) \partial_\nu \left[ \frac{1}{(u^t)^2} \right] = 0.
\end{equation}

We assume that $\varepsilon$ and $p$ are related by an equation of state, and introduce the baryonic (rest-mass) density $\rho_0$, as well as the specific enthalpy $h \coloneqq (\varepsilon + p)/\rho_0$, such that $\d p = \rho_0 \d h$. Equation \eqref{E-B_eq_rigid3} can be integrated by postulating that $h$ is a function of $u^t$: $h \eqqcolon \mathcal{H}(u^t)$. Observe that $\partial_\nu p = \rho_0 \partial_\nu h = \rho_0 \mathcal{H}' \partial_\nu u^t$, where $\mathcal{H}' \coloneqq \d h/\d u^t$. In this case, Eq.\ \eqref{E-B_eq_rigid3} can be written as
\begin{equation}
\rho_0 \mathcal{H}' \partial_\nu u^t  - \rho_0 \mathcal{H} \frac{1}{u^t}\partial_\nu u^t + 4 a^2 \left( \rho_0 \mathcal{H} + 2 p \right) \frac{1}{(u^t)^3} \partial_\nu u^t = 0,
 \end{equation}
and consequently
\begin{equation}
\rho_0 \mathcal{H}' - \rho_0 \mathcal{H} \frac{1}{u^t} + 4 a^2 \left( \rho_0 \mathcal{H} + 2 p \right) \frac{1}{(u^t)^3}   = 0.
\end{equation}
Dividing by $\rho_0$ (in the region filled with the fluid), we get
\begin{equation}\label{H_pol}
    \mathcal{H}' - \mathcal{H} \frac{1}{u^t} + 4 a^2 \left( \mathcal{H} + \frac{2 p}{\rho_0} \right) \frac{1}{(u^t)^3} = 0.
\end{equation}
Subsequent steps depend explicitly on the assumed equation of state.

\subsection{Constant energy density}

For $\varepsilon = \mathrm{const}$ (in the fluid region) we have $\d p/(\varepsilon + p) = \d\ln(\varepsilon + p)$. Since, on the other hand, $\d p/(\varepsilon + p) = \d \ln h$, it follows that $h$ is proportional to $\varepsilon + p$, or equivalently that $\rho_0 = \mathrm{const}$. It is then convenient to choose $\rho_0 = \varepsilon$ (this is also the choice made in \cite{Meinel2008}). Consequently,
\begin{equation}
\label{hconste}
    h = \frac{\varepsilon + p}{\varepsilon} = 1 + \frac{p}{\varepsilon},
\end{equation}
and
\begin{equation}
\label{pconste}
    \frac{p}{\rho_0} = \frac{p}{\varepsilon} = h - 1.
\end{equation}
Inserting this into Eq.\ \eqref{H_pol}, we obtain the equation
\begin{equation}\label{Heqconst}
    \mathcal{H}' -  \mathcal{H}\frac{1}{u^t} + 4a^2(3\mathcal{H}-2)\frac{1}{(u^t)^3} = 0,
\end{equation}
which, upon integration, yields
\begin{equation}\label{Heqconstsol}
    \mathcal{H}(u^t) =  -\frac{1}{18a}\left[\sqrt{6\pi} u^t \erf\left(\frac{\sqrt{6}a}{u^t}\right)\exp\left(\frac{6a^2}{(u^t)^2}\right) - 12a  
   \right] + \mathcal{C} u^t \exp\left(\frac{6a^2}{(u^t)^2}\right),
\end{equation}
where $\mathcal C$ is the integration constant. Using the property $\lim_{x \to 0}(\erf(x)/x) = 2/\sqrt{\pi}$, it can be checked that in the limit of $a \to 0$, one obtains
\begin{equation}\label{Heqsola0}
\mathcal H(u^t) = \mathcal C u^t \qquad (a = 0).
\end{equation}
Equation \eqref{Heqconstsol} can be treated as an integrated Euler--Bernoulli equation, replacing Eq.\ \eqref{Heqsola0} in the numerical scheme for solving the entire Einstein--Euler system of equations.

In the numerical implementation, we will also need the derivatives
\begin{subequations}
\begin{align}
    \frac{\partial \mathcal H}{\partial \mathcal C} &= u^t \exp\left(\frac{6 a^2}{(u^t)^2}\right), \\
    \frac{\partial \mathcal H}{\partial u^t} &= \mathcal{H} \frac{1}{u^t} - 4 a^2 (3 \mathcal{H} - 2) \frac{1}{(u^t)^3}, \label{Hderivut}
\end{align}
\end{subequations}
where Eq.\ \eqref{Hderivut} is a trivial consequence of Eq.\ \eqref{Heqconst}.

The behavior of higher derivatives of $\mathcal H$ with respect to $u^t$ has an impact on the convergence properties of the pseudospectral numerical scheme used in this work. Unfortunately, for $a \neq 0$ these derivatives seem to grow in the supremum norm, leading effectively to a decrease of the convergence of the entire numerical method. Table \ref{tab:derivatives1} gives sample values of the derivatives $\mathcal H^{(n)}(u_0^t) \coloneqq \partial^n \mathcal H(u_0^t)/\partial (u^t)^n$, where $u_0^t$ satisfies $\mathcal H(u_0^t) = 1$. This choice corresponds to the vanishing pressure, $p = 0$. In the examples collected in Tab.\ \ref{tab:derivatives1}, we set $\mathcal C = 1$ and choose the values of $a = 0.1$ and $a = 0.2$. In contrast to this behavior, for $a = 0$ one has the limit $\mathcal H (u^t) = \mathcal C u^t$ (Eq.\ \eqref{Heqsola0}). In this case $\mathcal H'(u^t) = \mathcal C$, and all higher derivatives vanish.

\begin{table}[t]
    \centering
    \caption{\label{tab:derivatives1} Derivatives of $\mathcal H$ with respect to $u^t$ for constant density models at $u^t = u^t_0$ such that $\mathcal H(u_0^t) = 1$ (this corresponds to $p = 0$). They can be computed directly from Eq.\ \eqref{Heqconst}, or from the explicit solution \eqref{Heqconstsol}.  We set $\mathcal C = 1$.}
    \begin{ruledtabular}
    \begin{tabular}{cccc}
     & $a = 0$, $u_0^t = 1$ & $a = 0.1$, $u_0^t = 0.965166$ & $a = 0.2$, $u_0^t = 0.833049$ \\
     \hline
     $\mathcal H^{(0)}(u_0^t)$ & 1 & 1 & 1 \\
     $\mathcal H^{(1)}(u_0^t)$ & 1 & $0.991602$ & $0.923646$ \\
     $\mathcal H^{(2)}(u_0^t)$ & 0 & $-0.0401569$ & $-0.102436$ \\
     $\mathcal H^{(3)}(u_0^t)$ & 0 & $0.308907$ & $1.4975$ \\
     $\mathcal H^{(4)}(u_0^t)$ & 0 & $-2.08438$ & $-13.8529$ \\
     $\mathcal H^{(5)}(u_0^t)$ & 0 & $15.2849$ & $136.129$ \\
     $\mathcal H^{(6)}(u_0^t)$ & 0 & $-125.617$ & $-1506.29$ \\
     $\mathcal H^{(7)}(u_0^t)$ & 0 & $1158.14$ & $18845.7$ \\
     $\mathcal H^{(8)}(u_0^t)$ & 0 & $-11906.4$ & $-264557.$ \\
     $\mathcal H^{(9)}(u_0^t)$ & 0 & $135467.$ & $4.12572 \times 10^6$ \\
     $\mathcal H^{(10)}(u_0^t)$ & 0 & $-1.6931 \times 10^6$ & $-7.07924 \times 10^7$
    \end{tabular}
    \end{ruledtabular}
\end{table}

\subsection{Polytropic fluids}

For the polytropic equation of state, $p = K \rho_0^\Gamma$, we have
\begin{equation}
\label{hforpoly}
h = 1 + \frac{\Gamma}{\Gamma - 1} \frac{p}{\rho_0},
\end{equation}
and consequently
\begin{equation}
\frac{p}{\rho_0} = \frac{\Gamma - 1}{\Gamma} (h - 1).
\end{equation}
Therefore, condition \eqref{H_pol} becomes
\begin{equation}
\mathcal{H}' -  \mathcal{H} \frac{1}{u^t} + 4 a^2 \left[ \mathcal{H} + 2 \frac{\Gamma - 1}{\Gamma} (\mathcal{H} - 1) \right]   \frac{1}{(u^t)^3}   = 0.
\end{equation}
The above equation can be integrated as
\begin{equation}
\label{HeqsolPoly}
    \mathcal{H}(u^t) =  - \frac{(\Gamma -1) \left[\sqrt{2\pi\Gamma} u^t \erf\left(\frac{a \sqrt{2(3\Gamma - 2)}}{\sqrt{\Gamma} u^t}\right)\exp\left(\frac{2 a^2 (3 \Gamma -2)}{\Gamma  (u^t)^2}\right) - 4a\sqrt{3\Gamma - 2}\right]}{2a 
   ( 3\Gamma - 2 )^{3/2}} + \mathcal{C} u^t \exp\left(\frac{2 a^2 (3 \Gamma -2)}{\Gamma  (u^t)^2}\right).
\end{equation}
Once again, the solution reduces to $\mathcal{H}(u^t) = \mathcal{C}u^t$ in the limit of $a \to 0$.

As before, in the numerical implementation, we will need the derivatives
\begin{subequations}
\begin{align}
    \frac{\partial \mathcal H}{\partial \mathcal C} &= u^t \exp\left(\frac{2 a^2 (3 \Gamma -2)}{\Gamma  (u^t)^2}\right), \\
    \frac{\partial \mathcal H}{\partial u^t} &= \mathcal{H} \frac{1}{u^t} - 4 a^2 \left[\mathcal{H} + 2 \frac{\Gamma - 1}{\Gamma} (\mathcal{H} - 1) \right]\frac{1}{(u^t)^3}.
\end{align}
\end{subequations}
Sample values of higher derivatives of $\mathcal H$ with respect to $u^t$ are collected in Tab.\ \ref{tab:derivatives2}. As in the constant-density case, they seem to grow in the supremum norm, impacting convergence properties of the numerical method used in this paper. For the polytropic equation of state, $p = K\rho_0^\Gamma$, these properties also depend on the assumed value of $\Gamma$. Note that Eq.\ \eqref{hforpoly} can be inverted, yielding
\begin{equation}
\label{ppoly}
    \rho_0 = \left[ \frac{h - 1}{(1 + n) K} \right]^n, \qquad p = K \left[ \frac{h - 1}{(1+n) K} \right]^{1+n},
\end{equation}
where $n \coloneqq 1/(\Gamma - 1)$ denotes the polytropic index. The expressions for $\rho_0$ and $p$ remain analytic functions of $h$ at $h = 1$ for $n = 0, 1,2,\dots$, but not for non-integer values of $n$. This impacts convergence properties of the pseudospectral AKM method, even in the unmagnetized case, corresponding to $a = 0$, as described in \cite{Meinel2008}.

Finally, let us notice that in both particular cases discussed above, the electric conductivity of the fluid actually vanishes. This follows from Eq.\ \eqref{sigma0} and the assumption that the specific enthalpy is a function of $u^t$. Indeed, if $u^t$ were constant on the part of spacetime filled with the fluid, then $h = \mathcal{H}(u^t)$ would be constant as well, which would lead to uniform distributions of thermodynamic quantities (e.g., the pressure). However, this turns out not to be the case for the numerical solutions discussed in Sec.\ \ref{Sec:NumSol}. Consequently, $u^t$ cannot be constant, and Eq.\ \eqref{sigma0} gives $F_{\rho\nu}u^{\nu} \neq 0$, $F_{\zeta\nu}u^\nu \neq 0$. As we already know that $\sigma F_{\mu\nu}u^\nu = 0$ (see the derivation of Eq.\ \eqref{j_decomp_sigma0}), we must have $\sigma = 0$.

\begin{table}[t]
    \centering
    \caption{\label{tab:derivatives2}Same as in Tab.\ \ref{tab:derivatives1}, but for the polytropic equation of state. We set $\mathcal C = 1$.}
    \begin{ruledtabular}
    \begin{tabular}{ccccc}
     & $\Gamma = 2$, $a = 0.1$,  & $\Gamma = 2$, $a = 0.2$,  & $\Gamma = 5/3$, $a = 0.1$, & $\Gamma = 5/3$, $a = 0.2$,\\
     & $u_0^t = 0.972324$ & $u_0^t = 0.872217$ & $u_0^t = 0.973741$ & $u_0^t = 0.879537$ \\
     \hline
     $\mathcal H^{(0)}(u_0^t)$ & $1$ & $1$ & $1$ & $1$\\
     $\mathcal H^{(1)}(u_0^t)$ & $0.98495$ & $0.905376$ & $0.983643$ & $0.901805$ \\
     $\mathcal H^{(2)}(u_0^t)$ & $0.00378691$ & $0.116285$ & $0.0122771$ & $0.153011$ \\
     $\mathcal H^{(3)}(u_0^t)$ & $0.0722495$ & $-0.0887735$ & $0.0273865$ & $-0.326636$ \\
     $\mathcal H^{(4)}(u_0^t)$ & $-0.648803$ & $-0.977188$ & $-0.382514$ & $0.735717$ \\
     $\mathcal H^{(5)}(u_0^t)$ & $5.20721$ & $13.9595$ & $3.38265$ & $0.388689$ \\
     $\mathcal H^{(6)}(u_0^t)$ & $-44.2599$ & $-161.221$ & $-29.9133$ & $-21.9982$ \\
     $\mathcal H^{(7)}(u_0^t)$ & $411.595$ & $1921.32$ & $283.591$ & $376.375$ \\
     $\mathcal H^{(8)}(u_0^t)$ & $-4210.37$ & $-24781.9$ & $-2928.97$ & $-5436.9$ \\
     $\mathcal H^{(9)}(u_0^t)$ & $47268.1$ & $349591.$ & $33021.2$ & $79576.8$ \\
     $\mathcal H^{(10)}(u_0^t)$ & $-579707$ & $-5.39059 \times 10^6$ & $-405334$ & $-1.23344 \times 10^6$
    \end{tabular}
    \end{ruledtabular}
\end{table}

\section{Electric charge}
\label{Sec:Charge}

The total electric charge $Q$ contained in a spacelike hypersurface $\Sigma$ is defined as the flux of the electric four-current through $\Sigma$:
\begin{equation}\label{Q_def}
    Q \coloneqq -\int_\Sigma j_\mu n^\mu \dd V.
\end{equation}
Here, $n^\mu$ denotes the future-directed unit vector normal to $\Sigma$, and $\dd V$ is the induced volume element on $\Sigma$. Taking as $\Sigma$ a hypersurface $t = \mathrm{const}$ (or its part containing the whole system), we obtain
\begin{equation}\label{n_cov}
    n_\mu = (-N,0,0,0),
\end{equation}
where $N = \sqrt{-g_{tt} + (g_{t \varphi})^2/g_{\varphi \varphi}}$ is the lapse function (see identity \eqref{g&gamma}). Consequently, $n_\mu j^\mu = j^t n_\mu k^\mu = -j^tN$, and
\begin{equation}\label{Q_jt}
    Q = \int_\Sigma j^t N \dd V.
\end{equation}
The charge associated with the electromagnetic field generated by a Killing vector $\xi^\mu$ will be further denoted as $Q_\xi$. For $\xi^\mu = a (k^\mu + \Omega \eta^\mu)$ (Eq.\ \eqref{Waldpot}) and $j^\mu$ satisfying \eqref{j_decomp_sigma0}, we have $j^t = \rho_\mathrm{c} u^t$, where  $\rho_\mathrm{c}$ is given by $\eqref{rhoc_rigid}$. In this case, Eq.\ \eqref{Q_jt} becomes
\begin{equation}\label{Q_fin}
    Q_\xi = -2a\int_\Sigma (\varepsilon + 3p) N \dd V.
\end{equation}
In particular, we obtain $Q_\xi < 0$ if $a > 0$, which is the case for all numerical examples presented below. Note also that the quantity appearing in the integrals \eqref{Q_def} and \eqref{Q_fin} can be interpreted as the electric charge density in the frame of a Zero Angular Momentum Observer (ZAMO):
\begin{equation}
\label{rho_ZAMO}
\rho_\mathrm{c}^\mathrm{(ZAMO)} \coloneqq -j^\mu n_\mu = -2a(\varepsilon + 3p)N.
\end{equation}

As a consistency check, let us reach the same result via a different route. To this end, we need to recall some standard definitions. The components of the Levi--Civita tensor (volume element) in a right-handed frame are given by
\begin{equation}
\label{epsilon}    \epsilon_{\mu\nu\rho\sigma} \coloneqq \sqrt{-g}[\mu,\nu,\rho,\sigma],
\end{equation}
where $g \coloneqq \det(g_{\mu\nu})$, and $[\mu,\nu,\rho,\sigma]$ denotes the completely antisymmetric symbol with $[0,1,2,3] \coloneqq 1$. The Hodge dual $\star\theta$  of a $p$-form $\theta$ is defined as the $(4 - p)$-form with the components
\begin{equation}
\star\theta_{\mu_1\ldots\mu_{4-p}} \coloneqq \frac{1}{p!}\theta_{\nu_1\ldots\nu_p}\epsilon\indices{^{\nu_1\ldots\nu_p}_{\mu_1\ldots\mu_{4-p}}}.
\end{equation}
Recalling an equivalent expression for the flux, we write Eq.\ \eqref{Q_def} as $Q = \int_\Sigma \star j$. By virtue of this formula, the inhomogeneous Maxwell equations \eqref{Maxwell} (recast in the form $\dd \star F = 4\pi \star j$, where $\dd$ is the exterior derivative operator), Stokes' theorem, and the relation $F = \dd\xi$, the total charge associated with $\xi^\mu$ can be computed as
\begin{equation}\label{Qtot}
    Q_\xi = \frac{1}{4\pi} \int_{\partial\Sigma} \star \d \xi = a(Q_k + \Omega Q_\eta).
\end{equation}
Here, $\partial\Sigma$ denotes the boundary of $\Sigma$, and the charges $Q_k$, $Q_\eta$ are related to the Komar mass $M_\mathrm{K} \coloneqq -(8\pi)^{-1}\int_{\partial\Sigma}\star\dd k$ and the angular momentum $J_\mathrm{K} \coloneqq (16\pi)^{-1}\int_{\partial\Sigma}\star\dd\eta$ of the system, respectively, by
\begin{equation}\label{qMJ}
Q_k = -2M_\mathrm{K}, \qquad Q_\eta = 4J_\mathrm{K}
\end{equation}
(see, e.g., Eqs.\ (8.186) and (8.187) of \cite{Straumann_textbook}).

Using identity \eqref{Killing_id} together with the Einstein equations, we write the Komar quantities as volume integrals over $\Sigma$ (see Eq.\ (11.2.10) of \cite{Wald_textbook})\footnote{Note that only the energy-momentum tensor of the fluid contributes to $M_\mathrm{K}$ and $J_\mathrm{K}$, as we do not include the electromagnetic part in the Einstein equations.}:
\begin{subequations}
\begin{align}
    M_\mathrm{K} &= 2\int_\Sigma\left(T^\mathrm{(FLUID)}_{\mu\nu} - \frac{1}{2}T^\mathrm{(FLUID)} g_{\mu\nu}\right)k^\mu n^\nu\dd V,\\
    J_\mathrm{K} &= -\int_\Sigma\left(T^\mathrm{(FLUID)}_{\mu\nu} - \frac{1}{2}T^\mathrm{(FLUID)} g_{\mu\nu}\right)\eta^\mu n^\nu\dd V.
\end{align}
\end{subequations}
By means of Eqs.\ \eqref{tmunufluid} and \eqref{Tfluid}, we find
\begin{equation}
T^\mathrm{(FLUID)}_{\mu\nu} - \frac{1}{2}T^\mathrm{(FLUID)} g_{\mu\nu} = (\varepsilon+p)u_\mu u_\nu + \frac{1}{2}(\varepsilon-p)g_{\mu\nu},
\end{equation}
and hence
\begin{subequations}
\begin{align}\label{M&J}
    M_{\rm K} &= 2\int_\Sigma\left[(\varepsilon+p)(u_\mu k^\mu)(u_\nu n^\nu) + \frac{1}{2}(\varepsilon-p)(k_\mu n^\mu)\right]\dd V,\\
    J_{\rm K} &= -\int_\Sigma(\varepsilon+p)(u_\mu \eta^\mu)(u_\nu n^\nu)\dd V,
\end{align}
\end{subequations}
where in the second expression we have used $\eta_\mu n^\mu=0$. Inserting the remaining scalar products,
\begin{equation}
    u_\mu k^\mu = u_t, \qquad u_\mu\eta^\mu = u_\varphi, \qquad n_\mu u^\mu = -Nu^t, \qquad n_\mu k^\mu = -N,
\end{equation}
into the integrals, and employing relations \eqref{qMJ}, we express Eq.\ \eqref{Qtot} as
\begin{equation}\label{Q_fin2}
\begin{aligned}
    Q_\xi = 2a(-M_\mathrm{K} + 2\Omega J_\mathrm{K}) &= 4a\int_\Sigma\biggl[(\varepsilon+p)u_tu^t + \frac{1}{2}(\varepsilon-p) + \Omega(\varepsilon+p)u_\varphi u^t\biggr]N\dd V\\
    &= 4a\int_\Sigma\biggl[(\varepsilon + p)(\underbrace{u_t u^t + u_\varphi u^\varphi}_{-1}) + \frac{1}{2}(\varepsilon - p)\biggr]N\dd V\\
    &= -2a\int_\Sigma(\varepsilon + 3p)N\dd V,
\end{aligned}
\end{equation}
which agrees with the result \eqref{Q_fin}. It should also be stressed that $Q_\xi$ is directly related to Komar quantities and the angular velocity $\Omega$ (the first equality in Eq.\ \eqref{Q_fin2}), whose values are provided by the AKM code as a part of the solution, and hence there is no need to calculate the integral of $(\varepsilon + 3p)N$ to find the total charge of a configuration.

\section{Numerical examples}
\label{Sec:NumSol}

Numerical examples given in this section were obtained by replacing the Euler--Bernoulli equation valid for non-magnetized configurations, essentially in the form \eqref{Heqsola0}, by one of the corresponding equations \eqref{Heqconstsol} or \eqref{HeqsolPoly} in the pseudospectral AKM code \cite{Ansorg2003,Meinel2008} for spheroidal configurations (rigidly rotating stars). Such a replacement implies a collection of other necessary changes in the code, which however do not interfere with the main spectral part encoding Einstein equations.

The Einstein equations are formulated assuming the metric \eqref{metricgen} in the form
\begin{equation}\label{metric}
    \d s^2 = - e^{2 \nu}\d t^2 + \rho^2 B^2 e^{-2 \nu} (\d\varphi - \omega\d t)^2 + e^{2 \mu}(\d \rho^2 + \d\zeta^2).
\end{equation}
The code solves elliptic equations for the functions $u$, $B$, $\omega$, and $\mu$, where $u := \nu - \ln B$, which for completeness we list in Appendix \ref{AppB}. The expressions associated with the fluid only appear in the source terms. They are obtained by first computing $u^t$ from Eq.\ \eqref{ut2}, then the specific enthalpy $h = \mathcal H(u^t)$ from Eqs.\ \eqref{Heqconstsol} or \eqref{HeqsolPoly}, and finally the pressure $p$ and the energy density $\varepsilon$, according to the assumed equation of state (Eqs.\ \eqref{pconste} and \eqref{ppoly}).

The numerical grid consists of two parts---one spanning the region within the star, and the other covering the outside region. In both cases the coordinates $(\rho,\zeta)$ are mapped to new coordinates $(s,t)$, covering rectangular regions $I^2 = [0,1] \times [0,1]$. All functions $u$, $B$, $\omega$, and $\mu$ are expanded in Chebyshev polynomials $T_j$, $j = 0,1,\dots,$ of $s$ and $t$, according to
\begin{equation}
    H^{(m)}(s,t) = \sum_{j,k=1}^m H^{(jk;m)}T_{j-1}(2s -1)T_{k-1}(2t - 1)
\end{equation}
(see Eq.\ (32) in \cite{Ansorg2003}). A number of additional equations correspond to the matching conditions at the surface of the star and the assumed values of the parameters. The precision of the method can be controlled mainly by specifying the number $m$, related to the number of polynomials used in the spectral expansion, and consequently the total number of expansion coefficients.

A set of resulting algebraic equations for the expansion coefficients is solved with usual Newton--Raphson iterations. As the initial data for this iterative procedure, we use solutions obtained in the non-magnetized case, corresponding to $a = 0$. Our solutions are then obtained in series characterized by increasing the value of $a$, where each solution in the series, except the last, serves as the initial data for the subsequent one.

Tables \ref{tab:HomogeneousVariousa}, \ref{tab:Poly1}, \ref{tab:Poly15} contain sample parameters characterizing the obtained solutions: the mass $M = M_\mathrm{K}$, the angular momentum $J = J_\mathrm{K}$, the angular velocity $\Omega$, and the ratio of the polar and equatorial radii $r_\mathrm{p}/r_\mathrm{e}$. All solutions reported in these tables were computed assuming $m = 40$.

\begin{figure}[t]
\centering
\includegraphics[width=0.5\linewidth]{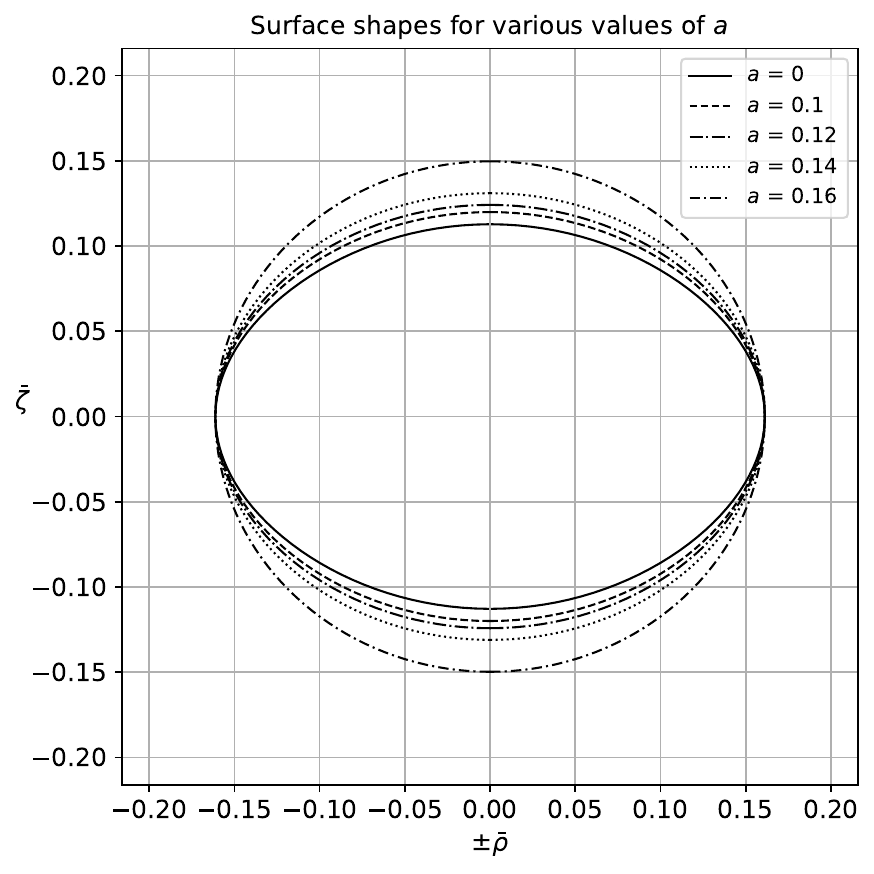} 
    \caption{Variation of the surface of the star with respect to $a$ for the homogeneous configuration characterized in Tab.\ \ref{tab:HomogeneousVariousa}. Here, $\bar{\rho} \coloneqq \rho\varepsilon^{1/2}$ and $\bar{\zeta} \coloneqq \zeta\varepsilon^{1/2}$ are dimensionless coordinates.}
    \label{fig:GeomStructure}  
\end{figure}

\begin{figure}[t]
    \centering   
    \begin{subfigure}{0.48\textwidth}
        \centering
        \includegraphics[width=\linewidth]{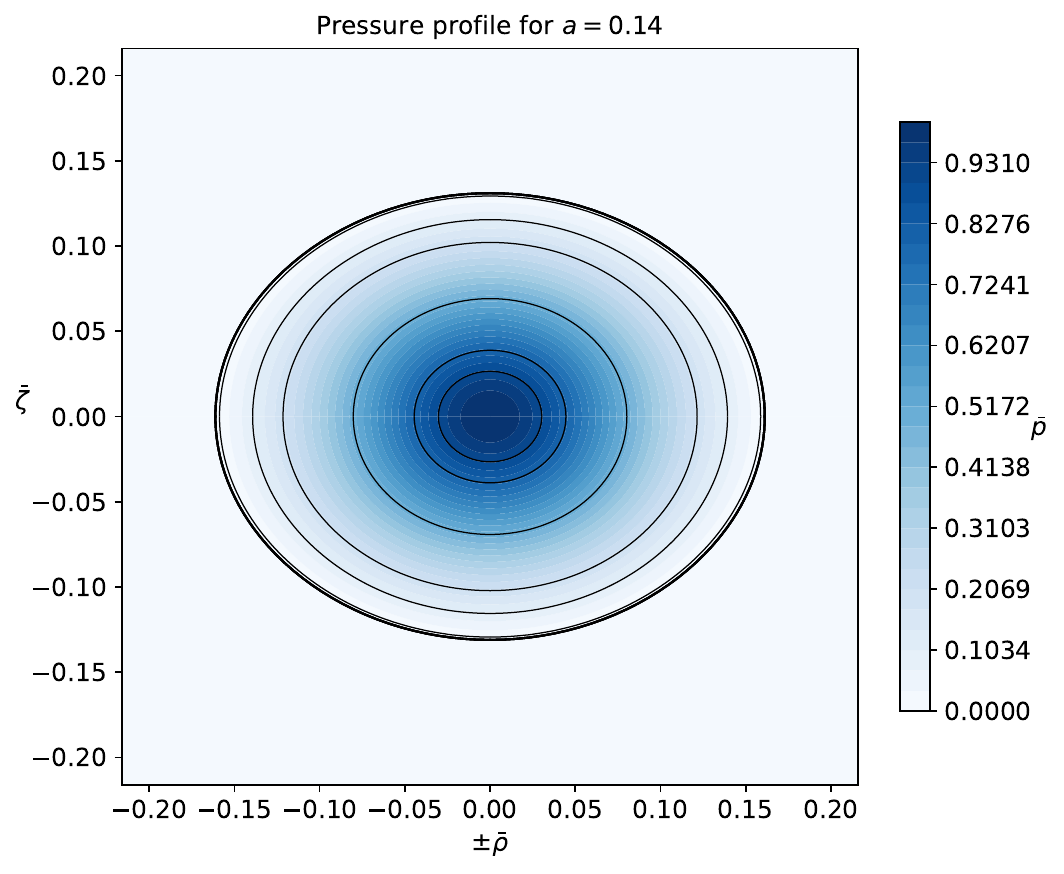}
    \end{subfigure}
    \hfill
    \begin{subfigure}{0.51\textwidth}
        \centering
        \includegraphics[width=\linewidth]{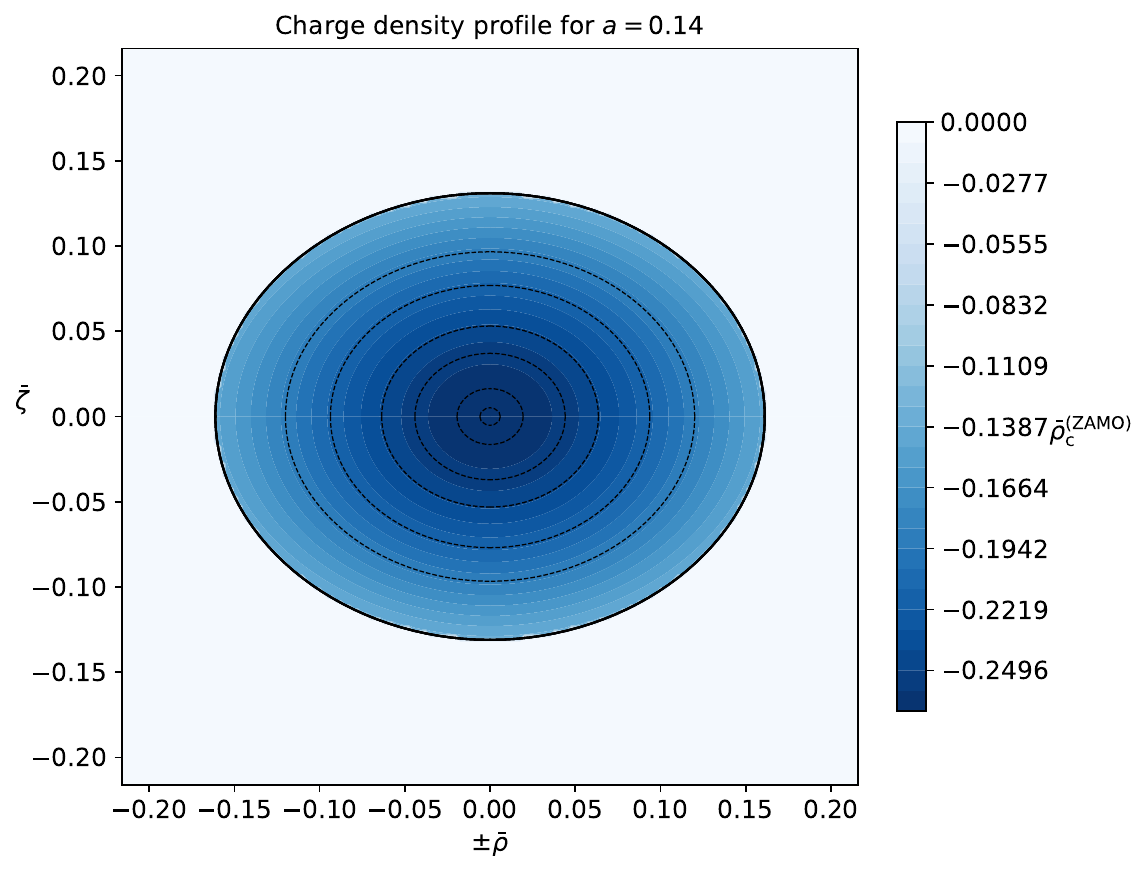}
    \end{subfigure}
    \caption{Distribution of the normalized pressure $\bar{p} \coloneqq p\varepsilon^{-1}$ (left) and electric charge density $\bar{\rho}_\mathrm{c}^{(\mathrm{ZAMO})} \coloneqq \rho_\mathrm{c}^{(\mathrm{ZAMO})}\varepsilon^{-1}$ (right) for the case $a = 0.14$ of the homogeneous configuration characterized in Tab.\ \ref{tab:HomogeneousVariousa}. For the meaning of $\bar{\rho}$ and $\bar{\zeta}$, see Fig.\ \ref{fig:GeomStructure}.}
    \label{fig:PressureAndChargeProfile}
\end{figure}

\begin{figure}[t]
    \centering
    \begin{subfigure}{0.48\textwidth}
        \centering
        \includegraphics[width=\linewidth]{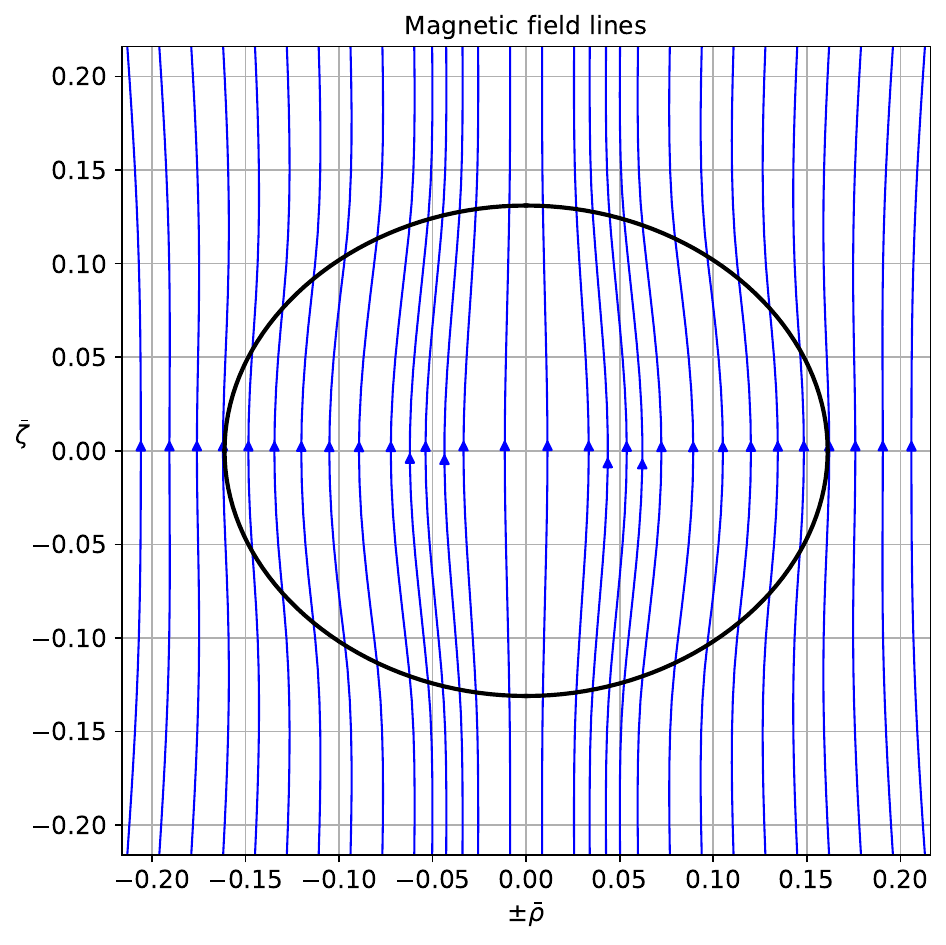}
        \label{fig:Bmag}
    \end{subfigure}
    \hfill
    \begin{subfigure}{0.48\textwidth}
        \centering
        \includegraphics[width=\linewidth]{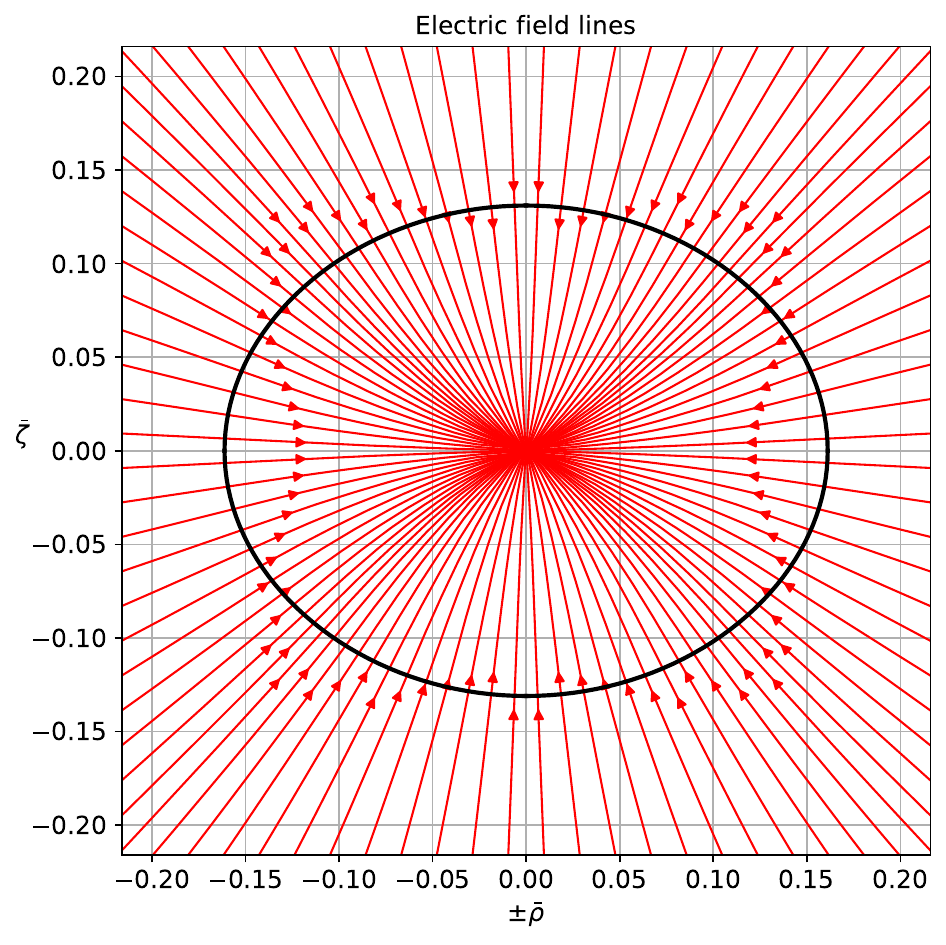}
        \label{fig:Elines}
    \end{subfigure}
    \caption{Magnetic (left) and electric (right) field lines, as seen by ZAMO observers, for the case $a = 0.14$ of the homogeneous configuration characterized in Tab.\ \ref{tab:HomogeneousVariousa}. For the meaning of $\bar{\rho}$ and $\bar{\zeta}$, see Fig.\ \ref{fig:GeomStructure}.}
    \label{fig:MagStructure}
\end{figure}

In Tab.\ \ref{tab:HomogeneousVariousa}, we collect parameters of a series of solutions with a constant energy density $\varepsilon$ (which is also used to establish a system of units). In this example, the equatorial radius is set to $r_\mathrm{e} \approx 0.161 \varepsilon^{-1/2}$, and the maximum value of the pressure (in the center of the star) is $p_\mathrm{c} = \varepsilon$. For a vanishing electromagnetic field ($a = 0$), these data correspond exactly to a solution reported in Tab.\ 1 of \cite{Ansorg2003}. We observe that by increasing slightly the value of $a$, one obtains solutions that become more and more prolate. The shapes of the surfaces of stars corresponding to different values of $a$ are depicted in Fig.\ \ref{fig:GeomStructure}. Moreover, Fig.\ \ref{fig:PressureAndChargeProfile} shows distributions within a star of the pressure (left panel) and the electric charge density in the ZAMO frame (right panel, see Eqs.\ \eqref{rho_ZAMO} and \eqref{rho_ZAMO_num}), obtained for $a = 0.14$ (note that, since both the electromagnetic potential and the stationary Killing vector are dimensionless, the coefficient $a$ is dimensionless as well, in view of Eq.\ \eqref{Waldpot}). Finally, Fig.\ \ref{fig:MagStructure} illustrates, for the same configuration, the magnetic and electric field lines with respect to ZAMO. They are computed according to the expressions collected in Appendix \ref{App:EMfields} (Eqs.\ \eqref{E} and \eqref{B}). As discussed in Sec.\ \ref{Sec:GenWaldMagn}, the magnetic field is constructed as asymptotically uniform and homogeneous, but it is nearly uniform and homogeneous also within the star. As expected, the electric field remains roughly spherically symmetric.

\begin{figure}[t]
    \centering
    \includegraphics[width=0.5\linewidth]{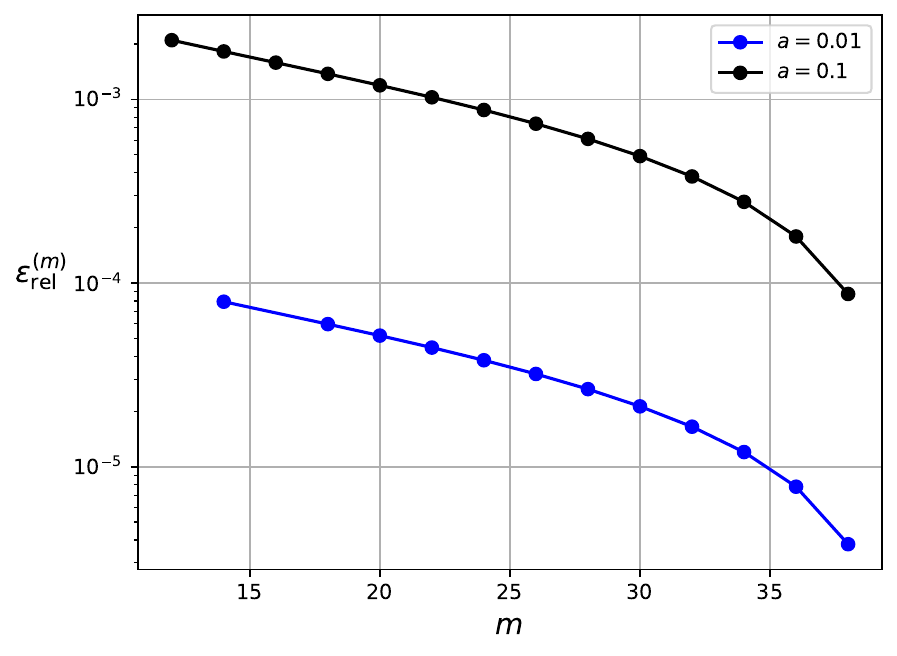}
    \caption{The relative deviation of the mass corresponding to Tab.\ \ref{tab:MagTab1AnsorgCombined}.}
    \label{fig:convergence}
\end{figure}

Results of sample convergence tests for constant energy density solutions with $a = 0.01$ and $a=0.1$ are provided in Tab.\ \ref{tab:MagTab1AnsorgCombined} and plotted in Fig.\ \ref{fig:convergence}. Here, for simplicity, we only consider the normalized mass $\bar M \coloneqq M \varepsilon^{1/2}$, and its deviation from the value computed with the maximum number of expansion polynomials, corresponding to $m = 40$. In both cases, the results converge with a growing $m$, although slower than for the non-magnetized solution with $a = 0$. In the latter case, sample convergence tests were described in \cite{Ansorg2003}. We emphasize that for $a \neq 0$, a relatively slow convergence is expected, given the behavior of the derivatives of $\mathcal H(u^t)$ summarized in Tab.\ \ref{tab:derivatives1}.

\begin{figure}[t]
    \centering
    \begin{subfigure}{0.48\textwidth}
        \centering
        \includegraphics[width=\linewidth]{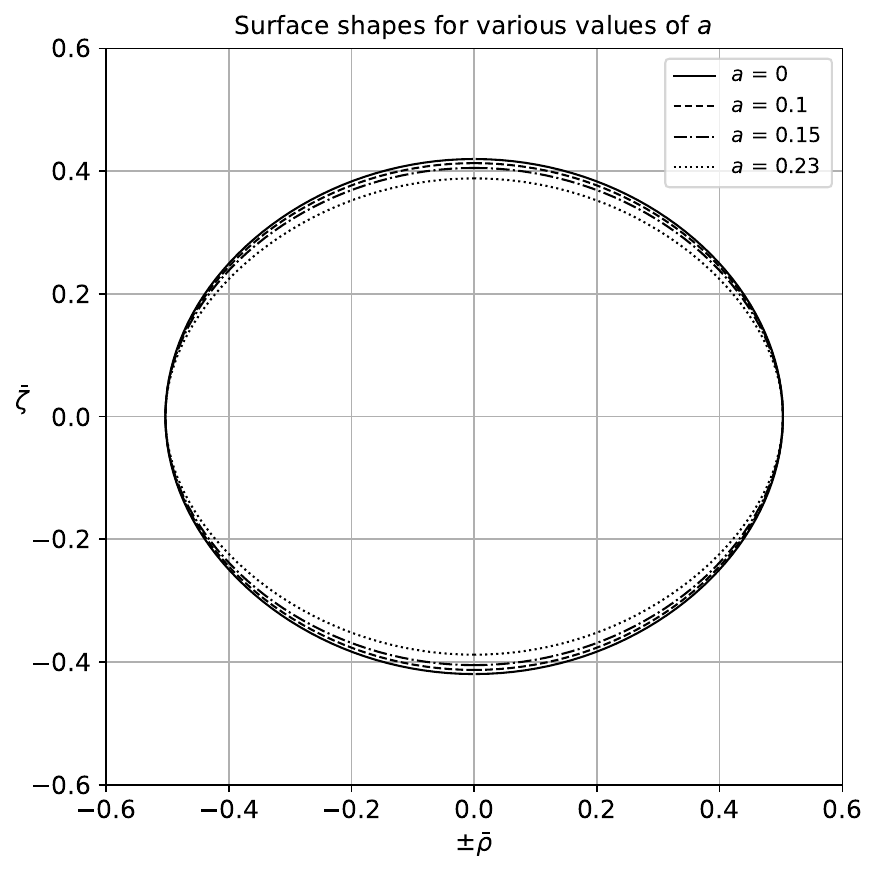}
        \label{fig:Poly1}
    \end{subfigure}
    \hfill
    \begin{subfigure}{0.48\textwidth}
        \centering
        \includegraphics[width=\linewidth]{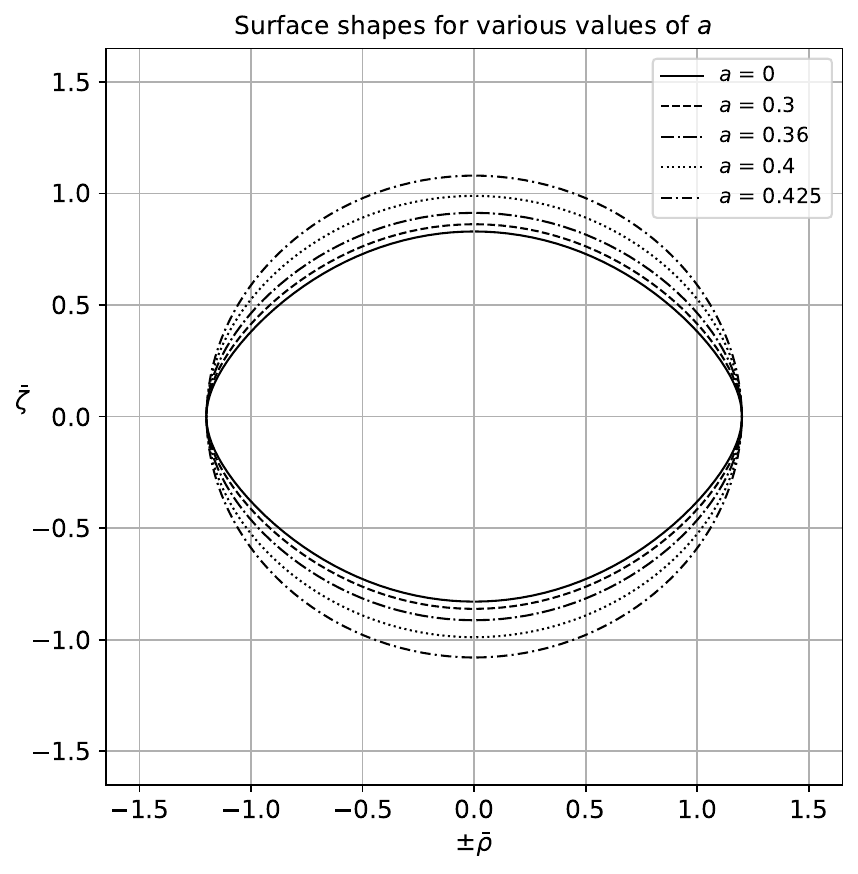}
        \label{fig:Poly15}
    \end{subfigure}  
    \caption{Left: Variation of the surface of the star with respect to $a$ for the polytropic configuration characterized in Tab.\ \ref{tab:Poly1} ($n = 1$). Here, $\bar{\rho} \coloneqq \rho K^{-1/2}$ and $\bar{\zeta} \coloneqq \zeta K^{-1/2}$ are dimensionless coordinates. Right: Same for the configuration from Tab.\ \ref{tab:Poly15} ($n = 1.5$). Here, $\bar{\rho} \coloneqq \rho K^{-3/4}$ and $\bar{\zeta} \coloneqq \zeta K^{-3/4}$.}
    \label{fig:GeomStructurePoly}
\end{figure}

Tables \ref{tab:Poly1} and \ref{tab:Poly15} present the solutions for configurations with the polytropic equation of state. Here, the system of units is provided by the constant $K$ (whose dimension is that of $(\text{length})^{2/n}$). In Tab.\ \ref{tab:Poly1}, the polytropic index is set to $n = 1$, the equatorial radius to $r_\mathrm{e} \approx 0.503K^{1/2}$, and the maximum value of the energy density (in the center of the star) to $\varepsilon_\mathrm{c} = K^{-1}$. For $a = 0$, these parameters correspond to a solution reported in Tab.\ 2 of \cite{Ansorg2003}. In Tab.\ \ref{tab:Poly15}, we provide data obtained for $n = 1.5$, $r_\mathrm{e} = 1.2K^{3/4}$, and $\varepsilon_\mathrm{c} = K^{-3/2}$. Perhaps the most obvious observation is that polytropic systems react differently to the presence of the electromagnetic field, depending on the value of the polytropic index: in the case of $n = 1$, the solutions with increasing $a$ become more and more oblate, while in the case of $n = 1.5$, they tend to be more prolate, similarly to constant energy density stars reported in Tab.\ \ref{tab:HomogeneousVariousa}. The shapes of the surfaces of stars for various values of $a$ are plotted in Fig.\ \ref{fig:GeomStructurePoly}.

\begin{table}[t]
    \centering
    \caption{Results for a homogeneous configuration with $m = 40$, $\bar{r}_\mathrm{e} \coloneqq r_\mathrm{e}\varepsilon^{1/2} =$ 1.611431133e-01, and $\bar{p}_\mathrm{c} \coloneqq p_\mathrm{c}\varepsilon^{-1} = 1$. The last two values are chosen to reproduce, for $a = 0$, the solution presented in Tab.\ 1 of \cite{Ansorg2003}. Here, $\bar{M} \coloneqq M\varepsilon^{1/2}$, $\bar{J} \coloneqq J\varepsilon$, and $\bar{\Omega} \coloneqq \Omega\varepsilon^{-1/2}$ are dimensionless quantities.}
    \label{tab:HomogeneousVariousa}
    \begin{ruledtabular}
    \begin{tabular}{ccccc}
        $a$ & $\bar{M}$ & $\bar{J}$ & $\bar{\Omega}$ & $r_\mathrm{p}/r_\mathrm{e}$ \\
        \hline
        0 & 1.357981788e-01 & 1.405859929e-02 & 1.411708483e+00 &  7.000000000e-01 \\
        0.05 & 1.344862249e-01 & 1.358138589e-02 & 1.394421527e+00 & 7.097751535e-01 \\
        0.1 & 1.301161811e-01 & 1.198058381e-02 & 1.326612863e+00 & 7.443937776e-01 \\
        0.11 & 1.287284270e-01 & 1.146597507e-02 & 1.301146026e+00 & 7.561510357e-01 \\
        0.12 & 1.270879008e-01 & 1.085153717e-02 & 1.268068198e+00 & 7.706002690e-01 \\
        0.13 & 1.251142658e-01 & 1.010070428e-02 & 1.223334021e+00 & 7.888601767e-01 \\
        0.14 & 1.226551418e-01 & 9.140335522e-03 & 1.158436380e+00 & 8.131375650e-01 \\
        0.15 & 1.193403428e-01 & 7.778403381e-03 & 1.049633453e+00 & 8.490527873e-01 \\
        0.16 & 1.131653339e-01 & 4.772140356e-03 & 7.276065309e-01 & 9.291739185e-01 \\
    \end{tabular}
    \end{ruledtabular}
\end{table}

\begin{table}[t]
    \centering
    \caption{Normalized mass $\bar{M}^{(m)} \coloneqq M^{(m)}\varepsilon^{1/2}$ and relative deviation $\varepsilon^{(m)}_\mathrm{rel} \coloneqq \abs{\bar{M}^{(m)} - \bar{M}^{(40)}}/\bar{M}^{(40)}$ for a homogeneous configuration with $r_\mathrm{p}/r_\mathrm{e} = 0.7$, $\bar{p}_c \coloneqq p_\mathrm{c}\varepsilon^{-1} = 1$, comparing 
    $a=0.01$ and $a=0.1$. 
    For $a=0.01$, $\bar{M}^{(40)} =$ 1.357467674e-01. 
    For $a=0.1$, $\bar{M}^{(40)} =$ 1.350949332e-01.}
    \label{tab:MagTab1AnsorgCombined}
    \begin{ruledtabular}
    \begin{tabular}{c cc cc}
    & \multicolumn{2}{c}{$a=0.01$} 
    & \multicolumn{2}{c}{$a=0.1$} \\
    \cline{2-3} \cline{4-5}
    $m$ 
    & $\bar{M}^{(m)}$ 
    & $\varepsilon^{(m)}_\mathrm{rel}$ 
    & $\bar{M}^{(m)}$ 
    & $\varepsilon^{(m)}_\mathrm{rel}$ \\
    \hline
    14 & 1.357575156e-01 & 7.917818092e-05 & 1.353412707e-01 & 1.823440271e-03 \\
    18 & 1.357548833e-01 & 5.978683542e-05 & 1.352810041e-01 & 1.377334484e-03 \\
    20 & 1.357537958e-01 & 5.177544081e-05 & 1.352560495e-01 & 1.192615966e-03 \\
    22 & 1.357528173e-01 & 4.456751705e-05 & 1.352336059e-01 & 1.026483892e-03 \\
    24 & 1.357519281e-01 & 3.801697144e-05 & 1.352132131e-01 & 8.755321526e-04 \\
    26 & 1.357511132e-01 & 3.201362160e-05 & 1.351945270e-01 & 7.372140555e-04 \\
    28 & 1.357503611e-01 & 2.647289646e-05 & 1.351772837e-01 & 6.095753312e-04 \\
    30 & 1.357496627e-01 & 2.132846064e-05 & 1.351612760e-01 & 4.910833008e-04 \\
    32 & 1.357490110e-01 & 1.652731588e-05 & 1.351463386e-01 & 3.805135751e-04 \\
    34 & 1.357484000e-01 & 1.202661812e-05 & 1.351323372e-01 & 2.768724642e-04 \\
    36 & 1.357478250e-01 & 7.790632433e-06 & 1.351191614e-01 & 1.793424711e-04 \\
    38 & 1.357472819e-01 & 3.790021004e-06 & 1.351067192e-01 & 8.724260748e-05 \\
    \end{tabular}
    \end{ruledtabular}
\end{table}

\begin{table}[t]
    \centering
    \caption{Results for a polytropic configuration with $m = 40$, $n = 1$, $\bar{r}_\mathrm{e} \coloneqq r_\mathrm{e}K^{-1/2} =$ 5.031368839e-01, and $\bar{\varepsilon}_\mathrm{c} \coloneqq \varepsilon_\mathrm{c}K = 1$. The last two values are chosen to reproduce, for $a = 0$, the solution presented in Tab.\ 2 of \cite{Ansorg2003}. Here, $\bar{M} \coloneqq MK^{-1/2}$, $\bar{J} \coloneqq JK^{-1}$, and $\bar{\Omega} \coloneqq \Omega K^{1/2}$ are dimensionless quantities. For this configuration, $\bar{p}_\mathrm{c} \coloneqq p_\mathrm{c}K =$ 3.819660113e-01.}
    \label{tab:Poly1}
    \begin{ruledtabular}
    \begin{tabular}{ccccc}
        $a$ & $\bar{M}$ & $\bar{J}$ & $\bar{\Omega}$ & $r_\mathrm{p}/r_\mathrm{e}$ \\
        \hline
        0 & 1.605611357e-01 & 9.491087857e-03 & 4.004385709e-01 &  8.340000000e-01 \\
        0.1 & 1.603004764e-01 & 9.836167453e-03 & 4.180256840e-01 & 8.207074524e-01 \\
        0.15 & 1.598650592e-01 & 1.019778288e-02 & 4.381235692e-01 & 8.051203930e-01 \\
        0.2 & 1.590603178e-01 & 1.058903708e-02 & 4.633686571e-01 & 7.849976567e-01 \\
        0.21 & 1.588400797e-01 & 1.066309578e-02 & 4.688668970e-01 & 7.805416475e-01 \\
        0.22 & 1.585965132e-01 & 1.073406246e-02 & 4.744818292e-01 & 7.759658487e-01 \\
        0.23 &  1.583279938e-01 & 1.080121203e-02 & 4.802004630e-01 & 7.712802685e-01 \\
        0.233 & 1.582423238e-01 & 1.082050475e-02 & 4.819344147e-01 & 7.698546552e-01 \\
    \end{tabular}
    \end{ruledtabular}
\end{table}

\begin{table}[t]
    \centering
    \caption{Results for a polytropic configuration with $m = 40$, $n = 1.5$, $\bar{r}_\mathrm{e} \coloneqq r_\mathrm{e}K^{-3/4} = 1.2$, and $\bar{\varepsilon}_\mathrm{c} \coloneqq \varepsilon_\mathrm{c}K^{3/2} = 1$. Here, $\bar{M} \coloneqq MK^{-3/4}$, $\bar{J} \coloneqq JK^{-3/2}$, and $\bar{\Omega} \coloneqq \Omega K^{3/4}$ are dimensionless quantities. For this configuration, $\bar{p}_\mathrm{c} \coloneqq p_\mathrm{c}K^{3/2} =$ 3.262363877e-01.}
    \label{tab:Poly15}
    \begin{ruledtabular}
    \begin{tabular}{ccccc}
        $a$ & $\bar{M}$ & $\bar{J}$ & $\bar{\Omega}$ & $r_\mathrm{p}/r_\mathrm{e}$ \\
        \hline
        0 & 1.965533724e-01 & 1.181134895e-02 & 2.268960283e-01 &  6.910957576e-01 \\
        0.05 & 1.963351437e-01 & 1.176974393e-02 & 2.268099219e-01 &  6.911742888e-01 \\
        0.1 & 1.956650566e-01 & 1.164053312e-02 & 2.264696655e-01 &  6.915671846e-01 \\
        0.15 & 1.945027769e-01 & 1.141088309e-02 & 2.256050700e-01 &  6.927988997e-01 \\
        0.2 & 1.928017673e-01 & 1.106018985e-02 & 2.236709336e-01 &  6.959480012e-01 \\
        0.25 & 1.905437619e-01 & 1.055899670e-02 & 2.196424382e-01 &  7.030861329e-01 \\
        0.3 & 1.877903002e-01 & 9.858001791e-03 & 2.115218253e-01 &  7.183171249e-01 \\
        0.35 & 1.847379077e-01 & 8.833383354e-03 & 1.949182909e-01 &  7.506099888e-01 \\
        0.36 & 1.841193047e-01 & 8.562229746e-03 & 1.897816405e-01 &  7.606945045e-01 \\
        0.37 & 1.835047783e-01 & 8.256191129e-03 & 1.837145423e-01 &  7.725971725e-01 \\
        0.38 & 1.828969347e-01 & 7.905110964e-03 & 1.764775104e-01 &  7.867379655e-01 \\
        0.39 & 1.822978003e-01 & 7.494079283e-03 & 1.677294410e-01 &  8.036795832e-01 \\
        0.4 & 1.817083691e-01 & 7.000195923e-03 & 1.569572898e-01 &  8.241990595e-01 \\
        0.41 & 1.811278063e-01 & 6.385857054e-03 & 1.433283080e-01 &  8.494105173e-01 \\
        0.42 & 1.805513218e-01 & 5.582310051e-03 & 1.253245016e-01 &  8.809922638e-01 \\
        0.425 & 1.802588671e-01 & 5.068670485e-03 & 1.137765197e-01 &  8.999615168e-01
    \end{tabular}
    \end{ruledtabular}
\end{table}

\section{Conclusions}
\label{Sec:Finis}

We have shown that the Wald solution---the electromagnetic field in the stationary and axially symmetric spacetime built from the available Killing vectors---can be compatible with non-vacuum spacetimes. In particular, it can be realized for rigidly rotating charged perfect fluids. In our examples, we considered configurations characterized by a constant energy density or polytropic fluids. In both cases, the equations expressing the conservation of the energy-momentum tensor can be integrated analytically, yielding an analog of the Euler--Bernoulli equation. As a result, the system can be described using the well-known frameworks developed for Einstein--Euler equations, with the only alterations present in the Euler--Bernoulli part.

The numerical examples provided in this paper were obtained by modifying the well-known (and publicly available) pseudospectral code by Ansorg, Kleinw\"{a}chter, and Meinel for solving Einstein--Euler equations describing spheroidal perfect-fluid configurations (rotating stars) \cite{Ansorg2003}. To maintain contact with the results of \cite{Ansorg2003}, we considered 3 series of solutions, branching from unmagnetized configurations of constant energy density or polytropes with polytropic indices $n = 1$ and $n = 1.5$. From a physical perspective, our solutions describe rigidly rotating (charged) stars embedded in an external uniform magnetic field. In particular, we found that the shape of a star may change in different ways when the electromagnetic field is ``added'' to the system: it becomes more and more spherical or flattened, depending on the equation of state and the polytropic index.

The solutions presented in this paper can also be regarded as a limit---for the vanishing electric conductivity $\sigma$---of a more general problem of constructing magnetohydrodynamical solutions representing rotating fluids immersed in asymptotically uniform magnetic fields (not necessarily described by Wald's magnetosphere).

\section{Acknowledgments}

We would like to thank Reinhard Meinel, Andreas Kleinw\"{a}chter, Gernot Neugebauer, and David Petroff for the kind permission to use the AKM code supplementing Ref.\ \cite{Meinel2008}. P.\ D.\ acknowledges support from the Faculty of Physics, Astronomy and Applied Computer Science under the Strategic Program Excellence Initiative at Jagiellonian University. A.\ T.\ acknowledges the Deutsche Forschungsgemeinschaft (DFG, German Research Foundation) funded under the project number 510727404. A.\ T.\ also acknowledges the Institute for Theoretical Physics of the Leibniz University of Hannover, for support and hospitality as a guest researcher. B.\ J.\ acknowledges grants SGS/24/2024 and IGS/27/2026 from the Silesian University in Opava and the Moravian-Silesian Region Foundation under the project ``Support for Talented Doctoral Students at the Silesian University in Opava 2022''.

\appendix

\section{Conservation of the energy-momentum tensor: fluid component}
\label{App:Bernoulli}

The standard relation for symmetric tensors gives
\begin{equation}
\nabla_\mu {T^\mathrm{(FLUID)}}\indices{^\mu_\nu} = \frac{1}{\sqrt{-g}} \partial_\mu \left( \sqrt{-g} {T^\mathrm{(FLUID)}}\indices{^\mu_\nu} \right) - \frac{1}{2} (\partial_\nu g_{\alpha \beta}) {T^\mathrm{(FLUID)}}^{\alpha \beta}.
\end{equation}
Recalling that $u^\rho = u^\zeta = 0$, while $u^t$, $u^\varphi$, $\varepsilon$, $p$ are functions of $(\rho,\zeta)$ only, and using the well-known identity $\partial_{\nu}g = gg^{\alpha\beta}\partial_{\nu}g_{\alpha\beta}$, one finds
\begin{equation}\label{divT}
    \begin{aligned}
        \nabla_\mu {T^\mathrm{(FLUID)}}\indices{^\mu_\nu} &= \frac{1}{\sqrt{-g}}\underbrace{\partial_\mu\left[\sqrt{-g}(\varepsilon + p)u^\mu u_\nu\right]}_0 + \underbrace{\frac{1}{\sqrt{-g}}\partial_\mu\left(\sqrt{-g}p\delta\indices{^\mu_\nu}\right)}_{\partial_\nu p + p\partial_\nu g/(2g)} - \frac{1}{2}(\partial_\nu g_{\alpha\beta})\left[(\varepsilon + p)u^\alpha u^\beta + pg^{\alpha\beta}\right]\\\
        &= \partial_\nu p + \frac{p}{2g}\underbrace{\left[\partial_\nu g - g(\partial_\nu g_{\alpha\beta})g^{\alpha\beta}\right]}_0 - \frac{1}{2}(\varepsilon + p)(\partial_\nu g_{\alpha\beta})u^\alpha u^\beta\\
        &= \partial_\nu p - \frac{1}{2}(\varepsilon + p)(\partial_\nu g_{\alpha\beta})u^\alpha u^\beta.
    \end{aligned}
\end{equation}
Now, taking the derivative of the normalization condition $g_{\alpha\beta}u^\alpha u^\beta = -1$, we obtain
\begin{equation}\label{guu}
    \begin{aligned}
        (\partial_\nu g_{\alpha\beta})u^\alpha u^\beta &= -g_{\alpha\beta}\partial_\nu(u^\alpha u^\beta)\\
        &= -g_{tt}\partial_\nu\left[(u^t)^2\right] - 2g_{t\varphi}\partial_\nu\left[\Omega(u^t)^2\right] - g_{\varphi\varphi}\partial_\nu\left[\Omega^2(u^t)^2\right]\\
        &= \underbrace{-(g_{tt} + 2g_{t\varphi}\Omega + g_{\varphi\varphi}\Omega^2)}_{(u^t)^{-2}}\partial_\nu\left[(u^t)^2\right] - (u^t)^2\left[2g_{t\varphi}\partial_\nu\Omega + g_{\varphi\varphi}\partial_\nu(\Omega^2)\right]\\
        &= \frac{2\partial_\nu u^t}{u^t} - 2u^t\underbrace{(g_{t\varphi}u^t + g_{\varphi\varphi}\Omega u^t)}_{u_\varphi}\partial_\nu\Omega\\
        &= 2\left(\frac{\partial_\nu u^t}{u^t} - u^t u_\varphi\partial_\nu\Omega\right).
    \end{aligned}
\end{equation}
Substituting this expression into Eq.\ \eqref{divT}, we finally arrive at Eq.\ \eqref{divT_fin}.

\section{Einstein equations}\label{AppB}

In this appendix we provide a few details of the numerical implementation, which in fact constitutes a small modification of the AKM code by Ansorg, Kleinw\"{a}chter, and Meinel \cite{Ansorg2003,Ansorg2005,Meinel2008}.

Equations come essentially from \cite{Bardeen1971}. The metric is assumed in the form \eqref{metric}. Three of Einstein equations can be written as (Eqs.\ (II.12), (II.13), and (II.14) of \cite{Bardeen1971})
\begin{subequations}
\begin{align}
    e^{-2 \mu} \left[ B^{-1} \nabla \cdot (B \nabla \nu) - \frac{1}{2} \rho^2 B^2 e^{-4 \nu} \nabla \omega \cdot \nabla \omega \right] &= 4 \pi \left[ (\varepsilon + p) \frac{1 + v^2}{1 - v^2} + 2 p \right], \label{Bardeen12} \\
    \frac{1}{2} \rho^{-1} B^{-2} e^{2 \nu - 2 \mu} \nabla \cdot \left( \rho^2 B^3 e^{-4 \nu} \nabla \omega \right) &= - 8 \pi (\varepsilon + p) \frac{v}{1 - v^2}, \label{Bardeen13} \\
    -\rho^{-1} B^{-1} e^{-2 \mu} \nabla \cdot \left( \rho \nabla B \right) &= - 16 \pi p. \label{Bardeen14}
\end{align}
\end{subequations}
Here
\begin{equation}
v \coloneqq (\Omega - \omega) \rho B e^{-2 \nu},
\end{equation}
and $\nabla$ denotes the flat gradient operator expressed in the cylindrical coordinates.

To obtain the equations used in the AKM code, we perform the following transformations: We replace function $\nu$ by $u \coloneqq \nu - \ln B$. Equation \eqref{Bardeen12}, multiplied by $e^{2\mu}$, gives
\begin{equation}
\label{AKMu}
\Delta u + B^{-1} \nabla B \cdot \nabla u + B^{-1} \Delta B - \frac{1}{2} \rho^2 B^{-2} e^{-4 u} \nabla \omega \cdot \nabla \omega = 4 \pi e^{2 \mu} \left[ (\varepsilon + p) \frac{1 + v^2}{1 - v^2} + 2 p \right].
\end{equation}
Equation \eqref{Bardeen14}, multiplied by $e^{2\mu}$, gives in turn
\begin{equation}
\label{AKMB}
- B^{-1} \Delta B - \rho^{-1} B^{-1} \nabla \rho \cdot \nabla B = - 16 \pi e^{2 \mu} p.
\end{equation}
Adding Eqs.\ \eqref{AKMu} and \eqref{AKMB}, one gets
\begin{equation}
\label{code0}
\Delta u + B^{-1} \nabla B \cdot \nabla u - 2 B^{-1} \partial_x B - \frac{1}{2} x B^{-2} e^{-4 u} \nabla \omega \cdot \nabla \omega = 4 \pi e^{2 \mu} \left[ (\varepsilon + p) \frac{1+v^2}{1-v^2} - 2 p \right],
\end{equation}
where we have introduced $x \coloneqq \rho^2$. Consequently $\nabla \rho \cdot \nabla f = \partial_\rho f = 2 \rho \partial_x f$, for any $f = B, \omega, \dots$ Equation \eqref{AKMB} now yields
\begin{equation}
\label{code1}
\Delta B + 2 \partial_x B = 16 \pi B e^{2 \mu} p.
\end{equation}
Next, multiply Eq.\ \eqref{Bardeen13} by $2 e^{2(\mu+u)}B\rho^{-1}$, to obtain
\begin{equation}
\label{code2}
\Delta \omega + 4 \partial_x \omega - B^{-1} \nabla B \cdot \nabla \omega - 4 \nabla u \cdot \nabla \omega = - 16 \pi (\varepsilon + p) e^{2 \mu} B e^{2 u} \rho^{-1} \frac{v}{1 - v^2}.
\end{equation}
The remaining equation is taken in the form (\cite{Ansorg2005}, Eq.\ (4d))
\begin{equation}
    \Delta_2 \mu - \frac{1}{\rho} \frac{\partial \nu}{\partial \rho} + \nabla \nu \cdot \nabla u - \frac{1}{4} \rho^2 B^2 e^{-4 \nu} \nabla \omega \cdot \nabla \omega = - 4 \pi e^{2 \mu} (\varepsilon + p).
\end{equation}
Here $\Delta_2 f \coloneqq \partial_\rho^2 f + \partial_\zeta^2 f$. Since $\Delta f = \frac{1}{\rho} \frac{\partial}{\partial \rho} \left( \rho \frac{\partial f}{\partial \rho} \right) + \frac{\partial^2 f}{\partial \zeta^2}$, one has $\Delta f = \Delta_2 f + \frac{1}{\rho} \frac{\partial f}{\partial \rho}$.
This finally gives
\begin{equation}
\label{code3}
\Delta \mu - 2 \partial_x \mu - 2 \partial_x u - 2 B^{-1} \partial_x B + \nabla u \cdot \nabla u + B^{-1} \nabla B \cdot \nabla u - \frac{1}{4} x B^{-2} e^{-4 u} \nabla \omega \cdot \nabla \omega = -4 \pi e^{2 \mu} (\varepsilon + p).
\end{equation} 

The AKM code solves Eqs.\ \eqref{code0}, \eqref{code1}, \eqref{code2}, and \eqref{code3} (and they appear in the code in that order).

\section{Electric and magnetic fields, and the charge density in the ZAMO frame}
\label{App:EMfields}

The electric and magnetic fields measured by an observer moving with the four-velocity $n^\mu$ are defined as $E_\mu \coloneqq F_{\mu\nu}n^\nu$ and $H_\mu \coloneqq -\star F_{\mu\nu}n^\nu$, or, equivalently,
\begin{equation}\label{B-1}
    E^\mu = g^{\mu\nu}g^{\rho\sigma}F_{\nu\rho}n_{\sigma}, \qquad H^\mu = -\frac{1}{2}\epsilon^{\mu\nu\rho\sigma}F_{\nu\rho}n_\sigma,
\end{equation}
where
\begin{equation}
    \epsilon^{\mu\nu\rho\sigma} = -\frac{1}{\sqrt{-g}}[\mu,\nu,\rho,\sigma],
\end{equation}
as follows from Eq.\ \eqref{epsilon}.

For a ZAMO observer, $n_\mu$ is of the form \eqref{n_cov}. Denoting $\gamma \coloneqq \det(g_{ij})$, and using the standard identity
\begin{equation}\label{g&gamma}
    \sqrt{-g} = N\sqrt{\gamma},
\end{equation}
we can therefore write Eqs.\ \eqref{B-1} as
\begin{equation}
    E^\mu = -Ng^{\mu\nu}g^{\rho0}F_{\nu\rho}, \qquad H^\mu = \frac{N}{2}\epsilon^{\mu\nu\rho0}F_{\nu\rho} = \frac{1}{2\sqrt{\gamma}}[0,\mu,\nu,\rho]F_{\nu\rho}.
\end{equation}
Hence, $E^0 = H^0 = 0$, and
\begin{equation}\label{B^i}
    E^i = -Ng^{i\nu}g^{\rho0}F_{\nu\rho}, \qquad H^i = \frac{1}{2\sqrt{\gamma}}[i,j,k]F_{jk} = \frac{1}{\sqrt{\gamma}}\sum_{j<k}[i,j,k]F_{jk}.
\end{equation}

In what follows, we work with metric functions $u$, $B$, $\omega$ and $\mu$, which are also computed in the AKM code, along with their derivatives over $x \coloneqq \rho^2$ and $y \coloneqq \zeta^2$. Since $e^u = B^{-1}e^\nu$, the metric \eqref{metric} takes the form
\begin{equation}\label{metric2}
    \d s^2 = - B^2 e^{2u}\d t^2 + \rho^2e^{-2u} (\d\varphi - \omega\d t)^2 + e^{2 \mu}(\d\rho^2 + \d\zeta^2),
\end{equation}
and the non-zero metric components can be expressed as
\begin{equation}
    g_{tt} = -B^2e^{2u} + \rho^2e^{-2u}\omega^2,\qquad g_{t\varphi} = -\rho^2e^{-2u}\omega, \qquad g_{\rho\rho} = g_{\zeta\zeta} = e^{2\mu}, \qquad g_{\varphi\varphi} = \rho^2e^{-2u}.
\end{equation}
It is then straightforward to calculate the determinants:
\begin{equation}\label{dets}
    \sqrt{-g} = B\rho e^{2\mu}, \qquad     \sqrt{\gamma} = \rho e^{-u + 2\mu},
\end{equation}
as well as the components of the inverse metric:
\begin{equation}\label{inv_g}
    g^{tt} = -B^{-2}e^{-2u}, \qquad g^{t\varphi} = -B^{-2}e^{-2u}\omega, \qquad g^{\rho\rho} = g^{\zeta\zeta} = e^{-2\mu}, \qquad g^{\varphi\varphi} = - B^{-2}e^{-2u}\omega^2 + \rho^{-2}e^{2u}.
\end{equation}
The lapse function $N$ may be found by combining Eqs.\ \eqref{g&gamma} and \eqref{dets}, which yields
\begin{equation}\label{N}
    N = Be^u.
\end{equation}

Moreover, given the potential \eqref{A_cov}, which can be written as
\begin{equation}\label{A}
    A_\mu = a(g_{\mu t} + \Omega g_{\mu\varphi}),
\end{equation}
we obtain
\begin{equation}\label{F}
    \begin{aligned}
        F_{t\rho} &= -\partial_\rho A_t, & F_{t\varphi} &= 0, & F_{t\zeta} &= -\partial_\zeta A_t,\\
        F_{\rho\varphi} &= \partial_\rho A_\varphi, & F_{\rho\zeta} &= 0, & F_{\varphi\zeta} &= -\partial_\zeta A_\varphi,
    \end{aligned}
\end{equation}
with the partial derivatives of
\begin{subequations}
\begin{align}
    A_t &= a\left(g_{tt} + \Omega g_{t\varphi}\right) = -a\left[B^2 e^{2u} + \rho^2e^{-2u}\omega\left(\Omega - \omega\right)\right],\\
    A_\varphi &= a\left(g_{t\varphi} + \Omega g_{\varphi\varphi}\right) = a\rho^2e^{-2u}\left(\Omega - \omega\right)
\end{align}
\end{subequations}
readily computable as:
\begin{subequations}\label{partial A}
\begin{align}
    \partial_\rho A_t &= -2a\rho\left\{2\left[B^2e^{2u} - \rho^2e^{-2u}\omega(\Omega - \omega)\right]\partial_x u + 2Be^{2u}\partial_x B + \rho^2e^{-2u}(\Omega - 2\omega)\partial_x \omega + e^{-2u}\omega(\Omega - \omega)\right\},\\
    \partial_\zeta A_t &= -2a\zeta\left\{2\left[B^2e^{2u} - \rho^2e^{-2u}\omega(\Omega - \omega)\right]\partial_y u + 2Be^{2u}\partial_y B + \rho^2e^{-2u}(\Omega - 2\omega)\partial_y \omega\right\},\\
    \partial_\rho A_\varphi &= -2a\rho e^{-2u}\left[2\rho^2(\Omega - \omega)\partial_x u + \rho^2\partial_x \omega - (\Omega - \omega)\right],\\
    \partial_\zeta A_\varphi &= -2a\rho^2\zeta e^{-2u}\left[2(\Omega - \omega)\partial_y u + \partial_y \omega\right].
\end{align}
\end{subequations}

Thanks to Eqs.\ \eqref{dets}, \eqref{inv_g}, \eqref{N}, \eqref{F}, and \eqref{partial A}, the electric and magnetic components \eqref{B^i} become, respectively,
\begin{subequations}\label{E}
\begin{align}
        E^\rho &= Ng^{\rho\rho}\left(g^{tt}F_{t\rho} - g^{t\varphi}F_{\rho\varphi}\right) = -2a\rho e^{u -2\mu}\left[2B\partial_x u + 2\partial_x B + \rho^2B^{-1}e^{-4u}(\Omega - \omega)\partial_x \omega\right],\\
        E^\varphi &= 0,\\
        E^\zeta &= Ng^{\zeta\zeta}\left(g^{tt}F_{t\zeta} + g^{t\varphi}F_{\varphi\zeta}\right) = -2a\zeta e^{u -2\mu}\left[2B\partial_y u + 2\partial_y B + \rho^2B^{-1}e^{-4u}(\Omega - \omega)\partial_y \omega\right],
\end{align}
\end{subequations}
and
\begin{subequations}\label{B}
\begin{align}
    H^\rho &= \frac{1}{\sqrt{\gamma}}F_{\varphi\zeta} = 2a\rho\zeta e^{-(u + 2\mu)}\left[2(\Omega - \omega)\partial_y u + \partial_y \omega\right],\\
    H^\varphi &= 0,\\
    H^\zeta &= \frac{1}{\sqrt{\gamma}}F_{\rho\varphi} = -2a e^{-(u + 2\mu)}\left[2\rho^2(\Omega - \omega)\partial_x u + \rho^2\partial_x \omega - (\Omega - \omega)\right].
\end{align}
\end{subequations}

For $\rho^2 + \zeta^2 \to \infty$, the metric \eqref{metric2} approaches the flat one in inertial cylindrical coordinates,
\begin{equation}
    \d s^2 = - \d t^2 + \d\rho^2 + \rho^2\d\varphi^2 + \d\zeta^2,
\end{equation}
which implies that $u,\omega,\mu \to 0$, $B \to 1$. Consequently, the components \eqref{E} and \eqref{B} simplify to
\begin{equation}
    E^i = 0, \qquad H^\rho = H^\varphi = 0, \qquad H^\zeta = 2a\Omega = 2b,
\end{equation}
i.e., to the uniform magnetic field associated with the potential \eqref{A} in Minkowski spacetime.

Finally, thanks to relation \eqref{N}, the electric charge density in a ZAMO frame, given by Eq.\ \eqref{rho_ZAMO}, can be expressed as
\begin{equation}\label{rho_ZAMO_num}
    \rho_\mathrm{c}^\mathrm{(ZAMO)} = -2a(\varepsilon + 3p)Be^u.
\end{equation}

\end{document}